\documentclass[fleqn,usenatbib]{mnras}

\usepackage[T1]{fontenc}

\DeclareRobustCommand{\VAN}[3]{#2}
\let\VANthebibliography\thebibliography
\def\thebibliography{\DeclareRobustCommand{\VAN}[3]{##3}\VANthebibliography}

\usepackage{graphicx}	% Including figure files
\usepackage{amsmath}	% Advanced maths commands
\usepackage{amssymb}	% Extra maths symbols
\usepackage{float}
\usepackage{lipsum}
\usepackage{xcolor}
\usepackage{natbib}
\usepackage{wasysym}
\usepackage{threeparttable}
\usepackage{epstopdf}
\usepackage{longtable}
\usepackage{rotating}
\usepackage{siunitx}

\DeclareSIUnit \parsec {pc}

\title[Evolution of dust]{Little evolution of dust emissivity in bright infrared galaxies from \boldmath{2 < \emph{z} < 6}}

\author[B.\,A.~Ward et al.]
{B.\,A.~Ward,$^{\! 1}$\thanks{E-mail: WardB2@cardiff.ac.uk}
S.\,A.~Eales,$^{\! 1}$
R.\,J.~Ivison$^{2,3,4,5}$ and
V.~Arumugam$^{6}$\\
\vspace*{1mm}\\
$^{1}$Cardiff Hub for Astrophysics Research and Technology (CHART), School of Physics and Astronomy, Cardiff University, The Parade, Cardiff CF24~3AA\\
$^{2}$European Southern Observatory (ESO), Karl-Schwarzschild-Strasse~2, D-85748 Garching, Germany\\
$^{3}$ARC Centre of Excellence for All Sky Astrophysics in 3 Dimensions (ASTRO 3D)\\
$^{4}$School of Cosmic Physics, Dublin Institute for Advanced Studies, 31 Fitzwilliam Place, Dublin D02 XF86, Ireland\\
$^{5}$Institute for Astronomy, University of Edinburgh, Royal Observatory, Blackford Hill, Edinburgh EH9 3HJ\\
$^{6}$Institut de Radioastronomie Millim\'{e}trique, 300 rue de la Piscine, Domaine Universitaire, 38406 Saint Martin d’H\`{e}res, France}

\date{Accepted XXX. Received YYY; in original form ZZZ}

\pubyear{2023}

\begin{document}
\label{firstpage}
\pagerange{\pageref{firstpage}--\pageref{lastpage}}
\maketitle

\begin{abstract}
Variations in the dust emissivity index, $\beta$, within and between galaxies, are evidence that the chemistry and physics of dust must vary on large scales, although the nature of the physical and/or chemical variations is still unknown. In this paper we estimate values of $\beta$ and dust temperature for a sample of 109 dusty star-forming galaxies (DSFGs) over the range, $2 < z < 6$. We compare the results obtained with both an optically-thin model and a general opacity model, finding that our estimates of $\beta$ are similar between the models but our estimates of dust temperature are not. We find no evidence of a change in $\beta$ with redshift, with a median value of $\beta = 1.96$ for the optically-thin model with a confidence interval (16--84\%) of 1.67 to 2.35 for the population. Using simulations, we estimate the measurement errors from our procedure and show that the variation of $\beta$ in the population results from intrinsic variations in the properties of the dust in DSFGs. At a fixed far-infrared luminosity, we find no evidence for a change in dust temperature, $T_{\textrm{dust}}$, with redshift. After allowing for the effects of correlated measurement errors, we find an inverse correlation between $\beta$ and $T_{\textrm{dust}}$ in DSFGs, for which there is also evidence in low-redshift galaxies.
\end{abstract}


%%%%%%%%%%%%%%%%%%%%%%%%%%%%%%%%%%%%%%%%%%%%%%%%%%

%%%%%%%%%%%%%%%%% BODY OF PAPER %%%%%%%%%%%%%%%%%%

\section{Introduction}
Despite contributing only a small percentage ($\lesssim 1$ per cent) of the mass of the interstellar medium (\citealt{Galliano_2018}), interstellar dust plays a pivotal role in the star-formation processes of galaxies. Dust efficiently absorbs optical and ultraviolet (UV) light from young OB-type stars and active galactic nuclei (AGN) and re-emits this radiation at longer wavelengths, in the far-infrared (FIR) to millimetre-wave (mm) regimes (e.g.~\citealt{Puget_1996}; \citealt{Dwek_1998}; \citealt{Fixsen_1998}; \citealt{Dole_2006}; \citealt{Driver_2008}). This dust obscuration hides star-formation activity in the most massive, dust-enshrouded galaxies in the early Universe. The most intense star formation occurs in dusty star-forming galaxies (DSFGs) at high redshifts ($z > 1$) where stars were produced at rates of $\gtrsim$ 100--1000\,$M_\odot$\,yr$^{-1}$, causing high infrared luminosities ($L_{\rm IR}$), often in excess of $\sim 10^{12}$\,L$_\odot$ (\citealt{Blain_2002}; \citealt{Casey_2014}; \citealt{Swinbank_2014}; \citealt{daCunha_2015, daCunha_2021}; \citealt{Dudzeviciute_2020}).

The first high-redshift DSFGs were discovered using the Sub-millimetre Common User Bolometer Array (SCUBA; \citealt{Holland_1999}) on the 15-m James Clerk Maxwell Telescope (JCMT) at 850\,\micron\ (\citealt{Smail_1997, Barger_1998, Hughes_1998, Eales_1999}). Since their discovery, many samples have been selected at wavelengths between 450\,\micron\ -- 2\,mm using SCUBA's successor, SCUBA-2 (\citealt{Holland_2013}), e.g.~ by \citet{Wang_2017} and \citet{Dudzeviciute_2020}, and by other single-dish FIR--mm facilities such as the \textit{Herschel Space Observatory} (e.g.~\citealt{Eales_2010}; \citealt{Oliver_2012}), AzTEC (e.g.~\citealt{Montana_2021}) and the South Pole Telescope (SPT; e.g.~\citealt{Vieira_2010}; \citealt{Everett_2020}), aided by the negative $K$ correction, which means that flux densities decline very slowly for a fixed $L_{\rm IR}$ at $z > 1$ (e.g.~\citealt{BL_1993}).

The very large star-formation rates in these galaxies suggest they contain very large masses of molecular gas. Since it is impossible to observe the molecular hydrogen directly, the mass of molecular gas in the DSFGs has been estimated by the standard method of making observations of a `tracer' of the hidden gas, then estimating the gas mass from the luminosity of the tracer and a calibration factor linking the two. The tracers that have been used are the CO molecule, dust grains and carbon atoms (e.g.~\citealt{Dunne_2022}). The dust method (e.g.~\citealt{Magdis_2012}; \citealt{Eales_2012}; \citealt{Scoville_2014}) is particularly useful for DSFGs because they are powerful sources of thermal continuum emission. Studies of the DSFGs and other high-redshift galaxies have shown that they contain a higher fraction of gas than observed in galaxies today (e.g.~\citealt{Tacconi_2010}; \citealt{Scoville_2016}; \citealt{Scoville_2017}; \citealt{Millard_2020}).

In considering this result, however, it should be remembered that the calibration factors are ultimately based on observations in our own Galaxy, and therefore the method is based on the implicit assumption that the physics and other properties of the tracer do not evolve with redshift. In the case of the dust method, this boils down to the assumptions that the gas-to-dust ratio and the physical and chemical properties of the dust do not change with redshift. Recent observational (e.g.~\citealt{Shapley_2020, Popping_2022}) and theoretical (e.g.~\citealt{Popping_2017, Li_2019}) studies have concluded that the dust-to-gas ratio does not appear to evolve with redshift.

One useful observational indicator of the physical and chemical properties of the dust grains is the dust emissivity spectral index, $\beta$, which describes the frequency dependence of the emissivity of the dust grains, with the optical depth of the dust being approximated as a power-law, $\tau \propto \nu^\beta$. Theoretical models of dust (e.g.~\citealt{Draine_1984}) predict that $\beta$ should range between 1--2 depending on the chemical composition of the dust grains. The value for dust in the Galaxy is remarkably uniform over the sky, with a value of $\beta = 1.51\pm 0.01$ (\citealt{Planck_2015}). Although the assumption necessary to use dust as a tracer of the hidden molecular hydrogen is that the properties of dust are the same in every galaxy, there is now plenty of evidence that this is not true. For example, \citet{Lamperti_2019} modelled the FIR SEDs of 192 nearby galaxies from JINGLES (the JCMT dust and gas In Nearby Galaxies Legacy Exploration Survey) and found $\beta$ values between 0.6 and 2.2. Also, there is now evidence for radial variation of $\beta$ within galaxies (e.g.~\citealt{Smith_2012, Draine_2014, Tabatabaei_2014, Hunt_2015, Whitworth_2019, Clark_2023}).

In this study, we test the assumption that the properties of dust are the same at all cosmic epochs by investigating whether there is evidence for evolution in the value of the dust emissivity index. We study bright FIR sources detected by \textit{Herschel} and SPT with spectroscopically determined redshifts between $2 \lesssim z_{\textrm{spec}} \lesssim 6$, and model their FIR dust emission to determine the value of $\beta$ for each galaxy. We investigate whether there is any difference between these values and those in galaxies today, and also whether there is any change over this redshift range, corresponding to the period between about a billion years after the Big Bang and the epoch at which cosmic star formation was at its peak (e.g.~\citealt{Lilly_1996}).

The paper is organised as follows. In Section \ref{sec:catalogue_creation} we describe the sample used during this analysis. In Section \ref{sec:sed_fitting} we derive the dust properties of our sources by fitting their SEDs with modified blackbodies. In Section \ref{sec:results} we consider the impact of the well-studied $\beta$--dust temperature degeneracy and the extent to which this biases our results. We also discuss the robustness of our estimated dust parameters by presenting simple simulations that imitate the SED fitting routine on mock \textit{Herschel} and SPT sources. Section \ref{sec:discussion} describes the results of our analysis and the suite of simulations used to predict the accuracy of our estimated dust properties. We also discuss the implications of our findings in the context of the existing literature for low redshift galaxies. Finally, we provide a summary of our results in Section \ref{sec:conclusions}. Throughout this paper we assume a $\Lambda$CDM cosmology with $H_0$ = 67.7\,km s$^{-1}$ Mpc$^{-1}$ and $\Omega_m$ = 0.31 (\citealt{Planck_2020}). 

\section{The sample}
\label{sec:catalogue_creation}

The sample studied here is a combination of DSFGs from {\it Herschel}, specifically from HerBS (\textit{Herschel} Bright Sources; \citealt{Bakx_2018}), and from the SPT (\citealt{Reuter_2020}).

\subsection{HerBS sub-sample}

The first sub-sample used in this study comes from the HerBS catalogue, a selection of the brightest high-redshift sources detected during the \textit{Herschel}-Astrophysical Terahertz Large Area Survey ({H-ATLAS; \citealt{Eales_2010}). The sample contains 209 galaxies with 500\,\micron\ flux densities above 80\,mJy and photometric redshift estimates in excess of 2.

Spectroscopic redshifts have been obtained for a selection of HerBS sources within the South Galactic Pole field of {\it H}-ATLAS, as part of the Bright Extragalactic ALMA Redshift Survey (BEARS; \citealt{Urquhart_2022}, \citealt{Bendo_2023}; \citealt{Hagimoto_2023}). These redshifts were measured from CO spectral lines detected with the band-3 receivers of the Atacama Compact Array (ACA), which is part of the Atacama Large Millimetre Array (ALMA), and also from spectral lines detected with the band-3 and -4 receivers of ALMA's 12-m array. The high spatial resolution of the ALMA band 4 images ($\sim 2"$) revealed that many HerBS sources are composed of more than one source; only half of the HerBS sources comprise a single source at ALMA's spatial resolution. In total, \citet{Urquhart_2022} measured 71 spectroscopic redshifts for 62 entries in the HerBS catalogue. Several HerBS sources comprise two or more ALMA sources with spectroscopic redshifts that agree well, suggesting either that the sources are physically associated galaxies or are multiple images caused by gravitational lensing. For this study, we have retained the HerBS sources which are multiples in the ALMA images, if the redshifts of the sources are within 0.1 of each other (we have taken the average redshift as that of the system). If there are multiple sources in the ALMA image within the {\it Herschel} beam but a spectroscopic redshift for only one source, we have assumed the sources are physical connected and used the redshift of the one source as the redshift of the system.

In our study we have used all HerBS sources with spectroscopic redshifts and photometry at 250, 350 and 500\,\micron\ (\textit{Herschel}-SPIRE), at 850\,\micron\ (\citealt{Bakx_2018}) and at 2 and 3\,mm from ALMA (\citealt{Bendo_2023}. We have also used photometry in ALMA band 6 (1.1--1.4\,mm) from the ALMARED survey (Jianhang Chen et al., in preparation) of potentially `ultrared' (see \citealt{Ivison_2016}) sources in {\it H}-ATLAS (project code 2018.1.00526.S). The minimum and maximum redshifts are $z_{\textrm{min}}$ = 1.569 and $z_{\textrm{max}}$ = 4.509.

\subsection{SPT sub-sample}
Our second sub-sample comprises the 81 DSFGs selected using the SPT (\citealt{Weiss_2013,Strandet_2016,Reuter_2020}). For a source to be included in this sub-sample, it must have a flux density above 25\,mJy at 870\,\micron. Redshifts for all SPT sources were obtained via blind searches targeting CO line emission (\citealt{Weiss_2013,Strandet_2016,Reuter_2020}). There are robust spectroscopic redshifts (from two or more CO lines) for all except two sources. We have retained the two because their spectroscopic redshifts are consistent with the redshifts estimated from photometry. The sub-sample has redshifts over the range 1.867 < $z_{\textrm{spec}}$ < 6.901 and a median redshift of $3.9\pm0.2$ (\citealt{Reuter_2020}). The SPT redshift distribution extends to higher redshifts than that of HerBS because the longer selection wavelength modifies the effect of the negative $K$ correction. The extreme apparent luminosities of the SPT sources implies that most of them are gravitationally lensed, with lensing models based on ALMA  observations giving a median lensing magnification factor of $5.5\times$ (\citealt{Spilker_2016}).

The SPT sources have photometry between 250\,\micron\ and 3\,mm, corresponding to a rest-frame range of 86\,\micron\ $\lesssim \lambda_{\textrm{rest}} \lesssim$ 380\,\micron, meaning that the peak of the dust emission ($\sim$ 100\,\micron) and Rayleigh-Jeans (R-J) tail are well sampled. The photometry includes flux densities measured at 250, 350 and 500\,\micron\ (\textit{Herschel}-SPIRE), at 870\,\micron\ (APEX-LABOCA), at 1.4 and 2\,mm (SPT) and at 3\,mm (ALMA). For 65 sources there are additional constraints at 100 and 160\,\micron\ from \textit{Herschel}-PACS. 

\subsection{Restrictions on sample for this study}

The two sub-samples contain 143 galaxies with spectroscopic redshifts. However, as we are interested in estimating the galaxy-integrated $\beta$ of each source, which is characterised by the slope of the dust emission on the R-J side of the Planck function, we restrict our final sample to those galaxies that have at least two photometric measurements at observed wavelengths beyond 1\,mm. This reduces our sample to a total of 109 galaxies (79 SPT DSFGs, plus 30 from HerBS/BEARS). While the fitting methods used in this work are the same for both sub-samples, we treat the two separately to look for effects that might have been caused by their different selection wavelengths and flux density limits. In Fig.~\ref{fig:spt_herbs_redshift} we show the redshift distributions of the two sub-samples. The photometry and redshifts of the HerBS-BEARS sources are tabulated in Appendix~\ref{tab:data_herbs}; the corresponding data used in this work for the SPT sources can be found in \citet{Reuter_2020}.

\begin{figure}
	\includegraphics[width=\columnwidth]{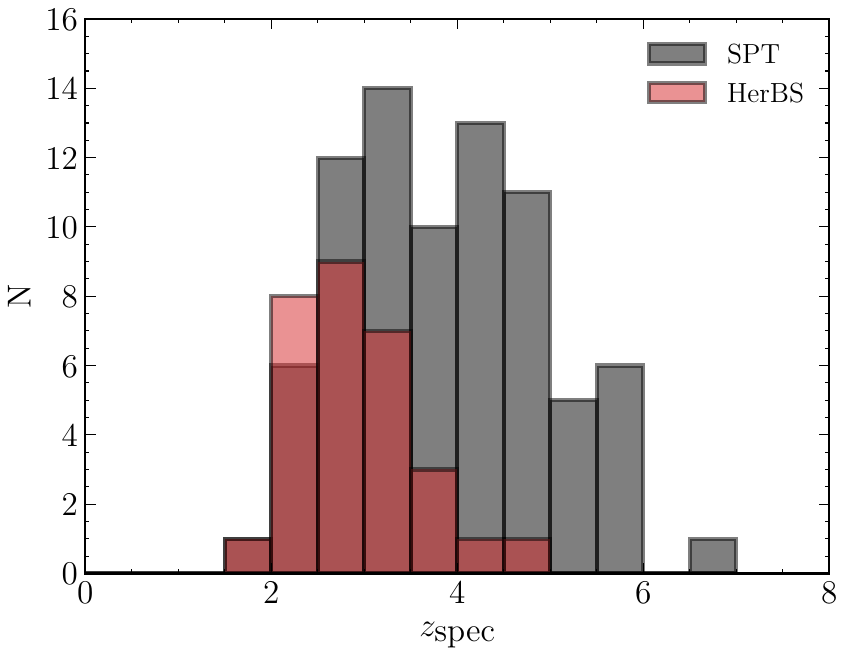}
	\caption{Spectroscopic redshift distributions for the SPT DSFGs (grey) and those from HerBS/BEARS (red).}
	\label{fig:spt_herbs_redshift}
\end{figure}

\section{SED fitting}
\label{sec:sed_fitting}

We model the FIR-to-millimetre spectra of our galaxies with a single modified blackbody (MBB) model combined with a mid-IR power-law, where the MBB represents the emission from the reservoir of cold dust, which represents most of the dust in the ISM, and the power-law represents emission from any dust in the vicinity of star-forming regions, young OB stars, or active galactic nuclei (AGN). The observed flux density can be expressed in the following form:

\begin{equation}
	S_{\nu, \textrm{obs}} =  
	\begin{cases}
	 	\frac{\Omega}{(1+z)^3} (1- e^{-\tau_\nu}) B_\nu(T_{\textrm{dust}}), & \nu \le \nu_{\rm c} \\
	 	 N \nu^{-\alpha}, & \nu > \nu_{\rm c} \\
	\end{cases}
\end{equation}

\noindent where $\Omega$ represents the solid angle subtended by the galaxy, $\tau_\nu$ is the dust optical depth, $B_\nu(T_{\textrm{dust}})$ is the Planck function assuming a characteristic dust temperature of the dust grains, $T_{\textrm{dust}}$, $N$ is a normalisation for the power-law which is tied to the normalisation of the blackbody, and $\alpha$ is the mid-IR power-law index. The value of $\nu_{\rm c}$ is the frequency at which the gradient of the MBB is equal to the value of $\alpha$.

We define the solid angle, $\Omega$, as $\frac{A(1+z)^4}{D_{\rm L}^2}$, where $A$ is the area of the source and $D_{\rm L}$ represents the luminosity distance. The optical depth, $\tau_\nu$, is defined as the product of dust surface mass density, $\Sigma_{\textrm{dust}} = M_{\textrm{dust}}/A$, and the dust opacity, $\kappa_\nu$, but is often assumed to take the form of a power-law, ($\nu/\nu_1)^\beta$, where $\nu_1$ represents the frequency at which the optical depth equals unity, marking the frequency of the transition between optically thin and optically thick dust. This parameter can be reformulated in terms of intrinsic properties by equating the two forms of $\tau$ such that $\nu_1 = \nu_0 (\kappa_0\Sigma_{\textrm{dust}})^\beta$. In a similar fashion, $\kappa_\nu$ may be assumed to take a power-law form: $\kappa_\nu = \kappa_0(\nu/\nu_0)^\beta$, where $\kappa_0$ is the emissivity of the dust grains per unit mass at a given reference frequency, $\nu_0$. In the following, we assume $\kappa_0$ = 0.077\,m$^2$\,kg$^{-1}$ at $\nu_0$ = 353\,GHz (\citealt{Dunne_2000, James_2003}). Combining these, we find the following equation for the observed flux density of a galaxy:

\begin{equation}
	S_{\nu, \textrm{obs}} =  
	\begin{cases}
		\frac{\mu A (1+z)}{D_{\rm L}^2} \Big(1- e^{-\frac{M_{\textrm{dust}}\kappa_\nu}{A}}\Big) B_\nu(T_{\textrm{dust}}), & \nu \le \nu_{\rm c} \\
		\mu N \nu^{-\alpha}, & \nu > \nu_{\rm c} \\
	\end{cases}
\end{equation}

\noindent where we have included a term, $\mu$, to account for the possibility of magnification due to gravitational lensing. For sources where there is no evidence of gravitational lensing, we assume $\mu = 1$.

As demonstrated by \citet{daCunha_2013}, the heating of dust by the cosmic microwave background (CMB) has a non-negligible effect on the shape of the FIR SED when the temperature of the CMB is a significant fraction of the dust temperature of a galaxy (see also \citealt{Zhang_2016} for the influence of the CMB on the measurement of structural and dynamical properties). Including the effects of the CMB on the observed dust continuum using the procedure outlined in \citet{daCunha_2013}, the CMB-adjusted MBB model is now given by: 

\begin{equation}
	S_{\nu, \textrm{obs}} =  
	\begin{cases}
		f_{\textrm{CMB}}\frac{\mu A (1+z)}{D_{\rm L}^2} \Big(1- e^{-\frac{M_{\textrm{dust}}\kappa_\nu}{A}}\Big) B_\nu(T_{\textrm{dust}}(z)), & \nu \le \nu_{\rm c} \\
		f_{\textrm{CMB}} \mu N \nu^{-\alpha}, & \nu > \nu_{\rm c} \\
	\end{cases}
\label{eq:mbb_general_opacity}
\end{equation}

\noindent where we have made the following two changes: i) the prefactor, $f_{\textrm{CMB}}$, denotes the fraction of the total dust emission that is observed against the background caused by the hotter CMB at higher redshifts, given by Equation~18 of \citet{daCunha_2013}, i.e.~$f_{\textrm{CMB}} = \frac{S_\nu^{\textrm{observed}}}{S_\nu^{\textrm{intrinsic}}} = 1 - \frac{B_\nu[T_{\textrm{CMB}}(z)]}{B_\nu[T_{\textrm{dust}}(z)]}$; and ii) we have re-defined the dust temperature to be a function of redshift due to the influence on dust grains of the warmer CMB at higher redshifts. The derived dust temperature is corrected for CMB effects using Equation~12 of \citet{daCunha_2013}, i.e.~$T_{\textrm{dust}}(z) = [T_{\textrm{dust}, 0}^{4+\beta} + T_{\textrm{CMB}, 0}^{4+\beta} ((1+z)^{4+\beta} - 1)]^{\frac{1}{4+\beta}}$, where $T_{\textrm{dust},0}$ is the dust temperature as observed at $z = 0$ and $T_{\textrm{CMB}, 0}$ = 2.73\,K is the CMB temperature at $z = 0$. Note that all further mention of the dust temperature refers to the CMB-corrected, luminosity-weighted dust temperature as given by the model described above, unless otherwise stated. Using this general form of the MBB, there are up to four free parameters: the dust mass, the dust temperature, $T_{\textrm{dust}}$, the dust emissivity index, $\beta$, and the frequency at which the dust opacity reaches unity. If the physical area of the region emitting the submillimetre emission is known, the opacity can be calculated from the surface density of the dust.

We  estimated the parameters of the dust using three different models. In the first model, we assume the dust is optically thin at all wavelengths ($\tau_\nu \ll 1$), which eliminates the fourth of the free parameters in the list above and modifies Eqn~\ref{eq:mbb_general_opacity} to:

\begin{equation}
	S_{\nu, \textrm{obs}} =  
	\begin{cases}
		f_{\textrm{CMB}}\frac{\mu (1+z)}{D_{\rm L}^2} M_{\textrm{dust}} \kappa_\nu B_\nu(T_{\textrm{dust}}(z)), & \nu \le \nu_{\rm c} \\
		f_{\textrm{CMB}}\mu N \nu^{-\alpha}, & \nu > \nu_{\rm c} \\
	\end{cases}
    \label{eq:modified_blackbody_optically_thin}
\end{equation}

However, since in DSFGs a large amount of dust is packed into a small region (e.g.~\citealt{Ikarashi_2017}), it is possible that the dust is optically thick at one or more of the wavelengths covered by our photometry (e.g.~\citealt{Conley_2011, Casey_2019, Cortzen_2020}). We therefore also used two models in which the dust becomes optically thick at wavelengths below $\lambda_1$ = 100\,\micron\ or 200\,\micron.

Some of the SPT sources have measurements of the sizes of the submillimetre-emitting region, typically $\sim 1$\,kpc (\citealt{Spilker_2016}). For this subset of sources, we also fitted a modified MBB using the size of the emitting region to calculate the opacity directly.

To estimate the mass and luminosity of the dust we require an estimate of the lensing magnifications, which we have taken to be those quoted in the lens modelling of \citet{Spilker_2016} for the SPT sources and from the correlation between CO luminosity and line width as determined by \citet{Urquhart_2022} for HerBS, building on the work of \citet{Harris_2012}. During SED fitting we assume flat priors on all free parameters and take the best-fitting SED to be the median value of the posterior distribution for each parameter. The (1-$\sigma$) uncertainties are quoted at the 16th and 84th percentiles. Calibration errors are added in quadrature with the flux density uncertainties assuming absolute calibrations of 7 per cent (\textit{Herschel}-PACS), 5.5 per cent (\textit{Herschel}-SPIRE), 5 per cent (SCUBA-2), 12 per cent (APEX-LABOCA), 7 per cent (SPT) and 10 per cent (ALMA). 

\section{Results}
\label{sec:results}

\subsection{Comparison between the optically thin and general opacity models}
\label{sec:comparison_optically_thin_and_general_opacity}

Fig.~\ref{fig:stacked_posteriors} shows the stacked posterior probability distributions for the dust parameters derived for the two samples. The median values and the 16th and 84th percentiles of the probability distributions are listed in Table~\ref{tab:parameter_results}. The median value  of the main parameter of interest, the dust emissivity spectral index, covers only a range of 0.2 for the three models, showing that it doesn't make much difference what one assumes about the opacity, a conclusion also reached in a study of a different sample of DSFGs by \citet{Ismail_2023}. We note, however, that the general opacity model with the highest value of $\lambda_1$ does give the highest value of $\beta$, as observed previously by \citet{McKay_2023}. Using a sample of 870\,\micron-selected galaxies in GOODS-S, \citet{McKay_2023} found good agreement in the median $\beta$ between an optically thin model (1.78) and an MBB model with $\lambda_1 = 100$\,\micron\ (1.80), but this increased to 2.02 for a model with $\lambda_1 = 200$\,\micron. Some recent studies have suggested that $\beta > 2$ is common among high-redshift galaxies (\citealt{Casey_2019}; \citealt{Casey_2021}; \citealt{Cooper_2022}), but this may be due to these studies using a higher value for the wavelength at which the dust becomes optically thick. 
  
\begin{figure*}
	\centering
	\includegraphics[width=\textwidth]{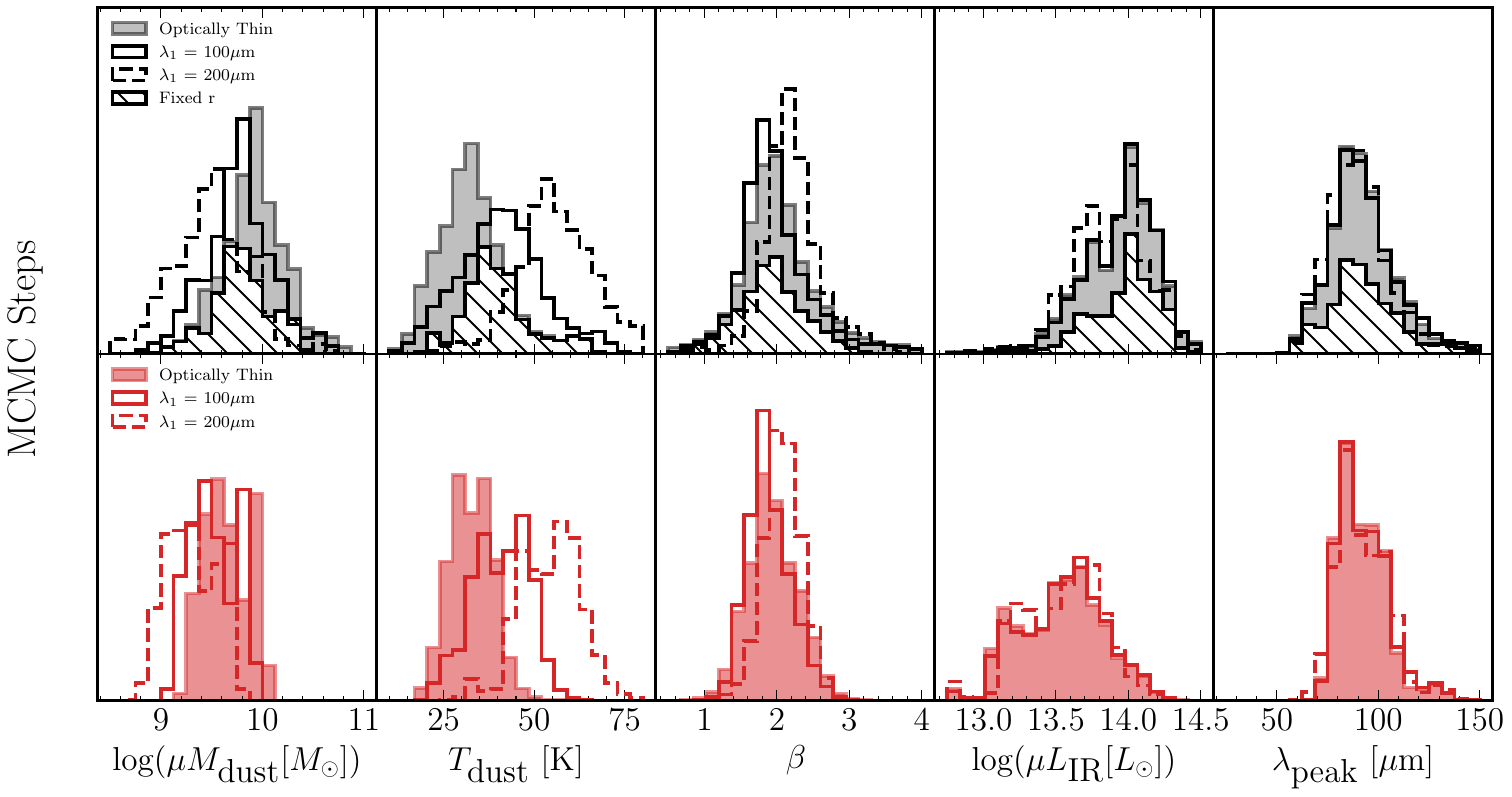}
	\caption{Stacked posterior distributions of log($\mu M_{\textrm{dust}}$), $T_{\textrm{dust}}$, $\beta$, log($\mu L_{\textrm{IR}}$) (rest-frame 8--1000\,\micron) and $\lambda_{\textrm{peak}}$ for SPT (top panels) and HerBS galaxies (bottom panels). The posterior distribution for each MBB model is illustrated as follows: optically thin (shaded); $\lambda_1 = 100$\,\micron\ (solid line); $\lambda_1 = 200$\,\micron\ (dashed line); fixed continuum size (hatched).}
	\label{fig:stacked_posteriors}
\end{figure*}

\begin{table}
    \caption{Median and 1$\sigma$ errors, estimated from the 16th, 50th and 84th percentiles of the stacked posterior distribution, for the parameters presented in Fig.~\ref{fig:stacked_posteriors}.}
    \centering
    \begin{tabular}{p{2.5cm}p{1.5cm}p{1.5cm}p{1.5cm}}
        \hline
        \hline
	Parameter & Optically& $\lambda_1 =$ & $\lambda_1 =$ \\
                  &thin&100\,\micron &200\,\micron\\
\hline
		 & \multicolumn{3}{c}{SPT} \\
        
        log($\mu M_{\textrm{dust}} [$M$_\odot]$) & $9.92_{-0.31}^{+0.30}$ & $9.77_{-0.34}^{+0.31}$ & $9.49_{-0.41}^{+0.37}$ \\
		$T_{\textrm{dust}}$ [K] & $32.07_{-8.44}^{+8.43}$ & $40.77_{-11.77}^{+10.90}$ & $55.63_{-9.28}^{+11.20}$ \\
		$\beta$ & $1.98_{-0.39}^{+0.54}$ & $1.91_{-0.32}^{+0.50}$ & $2.18_{-0.30}^{+0.38}$ \\
		log($\mu L_{\textrm{IR}} [$L$_\odot]$) & $13.98_{-0.31}^{+0.19}$ & $13.99_{-0.30}^{+0.19}$ & $13.89_{-0.27}^{+0.23}$ \\
		$\lambda_{\textrm{peak}}$ [\micron] & $90.70_{-12.72}^{+17.33}$ & $90.70_{-13.02}^{+18.36}$ & $89.28_{-13.39}^{+15.21}$ \\
		
		\hline
		& \multicolumn{3}{c}{HerBS} \\
        
        log($\mu M_{\textrm{dust}} [$M$_\odot]$) & $9.62_{-0.22}^{+0.30}$ & $9.49_{-0.22}^{+0.31}$ & $9.30_{-0.26}^{+0.33}$ \\
		$T_{\textrm{dust}}$ [K] & $32.73_{-5.92}^{+5.85}$ & $41.16_{-8.20}^{+7.48}$ & $55.18_{-9.23}^{+7.25}$ \\
		$\beta$ & $1.92_{-0.32}^{+0.39}$ & $1.85_{-0.25}^{+0.34}$ & $2.08_{-0.25}^{+0.26}$ \\
		log($\mu L_{\textrm{IR}} [$L$_\odot]$) & $13.55_{-0.37}^{+0.27}$ & $13.57_{-0.37}^{+0.26}$ & $13.57_{-0.34}^{+0.25}$ \\
		$\lambda_{\textrm{peak}}$ [\micron] & $90.03_{-9.28}^{+12.70}$ & $90.36_{-9.46}^{+13.42}$ & $89.96_{-9.46}^{+14.92}$ \\
        \hline
    \end{tabular}
    \label{tab:parameter_results}
\end{table}

The assumption that the dust is optically thin does, however, have a much larger effect for some of the other parameters. There is a clear trend to higher median dust temperatures with increasing $\lambda_1$, with increases of $\sim$ 10\,K and $\sim$ 20\,K from the optically-thin model to the $\lambda_1 = 100$\,\micron\ and to the $\lambda_1 = 200$\,\micron\ general opacity models, respectively. The dust masses decrease by approximately 0.1\,log($\mu $M$_\odot$) and 0.4\,log($\mu $M$_\odot$) from the optically-thin model to the $\lambda_1 = 100$\,\micron\ and the $\lambda_1 = 200$\,\micron\ general opacity models, respectively, though we note that these decreases are smaller than the 1$\sigma$ widths of the probability distributions. There is no significant difference in IR luminosities or peak wavelengths between the models.

Fig.~\ref{fig:stacked_posteriors} also shows the posterior probability distributions for the subset of SPT galaxies for which we are able to estimate the area of the dust region (hatched histograms). This distribution agrees well with the probability distributions for the optically-thin model and for the general-opacity model with $\lambda_1 = 100$\,\micron, but not for the one with $\lambda_1 = 200$\,\micron. For this subset of galaxies we find a median value of  $\lambda_1 = 88$\,\micron\ (16th and 84th percentile range 44--224\,\micron). However, we note that this value is inversely proportional to the value assumed for the dust opacity ($\kappa_{\nu}$) and since the opacity of interstellar dust is still very uncertain (\citealt{Clark_2016}), the frequency at which the opacity is unity is also uncertain.

Fig.~\ref{fig:comparison_optically_thin_general_opacity} shows the median values of the parameters estimated from the optically-thin model versus those estimated from the general-opacity models. As we discussed above, the estimates from the two types of model agree well for $\beta$, $\lambda_{\textrm{peak}}$ and far-infrared luminosity, but there are systematic difference for $T_{\textrm{dust}}$ and $M_{\textrm{dust}}$. The probability distributions for the optically-thin model are a good match to those for the subset of galaxies for which we know the size of the dust region, and therefore the parameter estimates we discuss in the rest of this paper have been derived using the optically-thin model. The panels in Fig.~\ref{fig:comparison_optically_thin_general_opacity} show the size of the systematic errors on these estimates if the dust is actually optically thick.

\begin{figure*}
	\centering
	\includegraphics[width=0.55\textwidth, height=0.96\textheight]{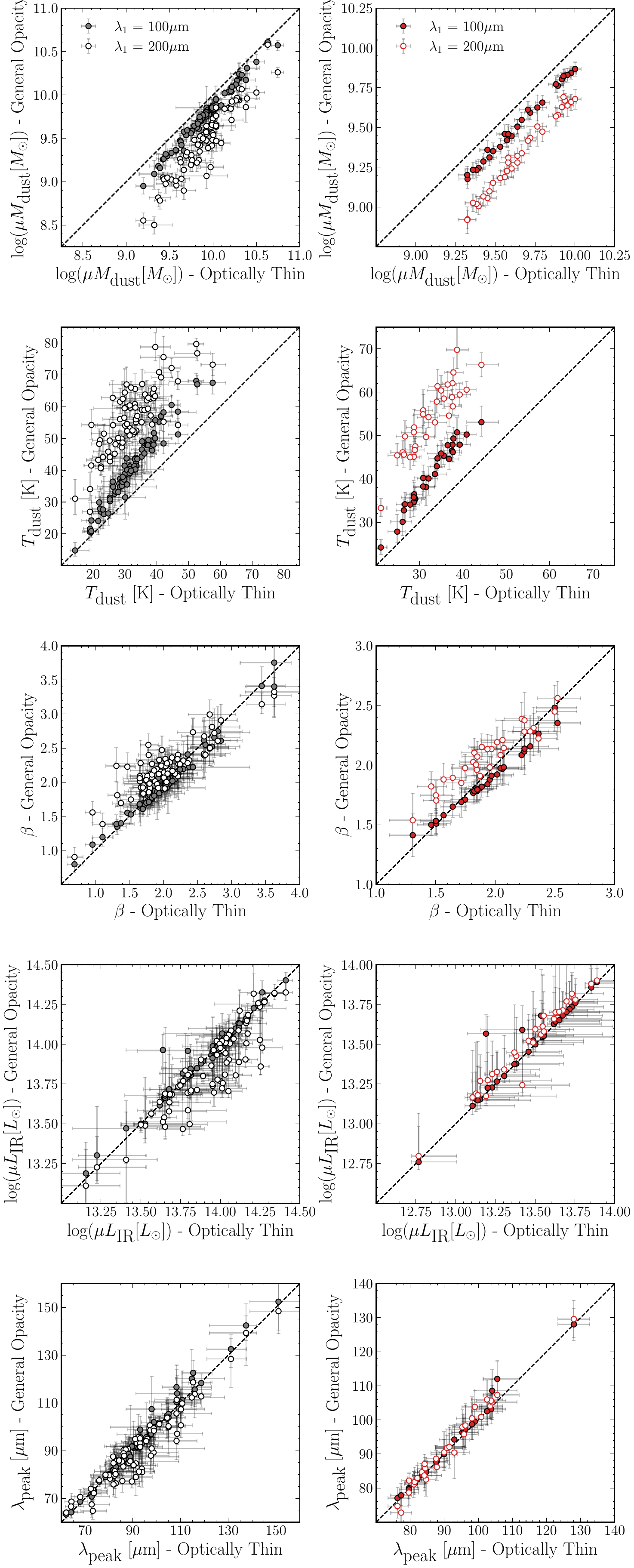}
	\caption{Comparison between the parameters derived from the optically-thin and general-opacity models for the galaxies in the SPT (left column) and HerBS (right column) sub-samples. The general opacity models are shown as filled and open circles for $\lambda_1$ = 100\,\micron\ and 200\,\micron, respectively.}
	\label{fig:comparison_optically_thin_general_opacity}
\end{figure*}

\subsection{The $\beta$--dust temperature relationship}
\label{sec:beta_t_correlation}

Where an MBB is fitted to an empirical SED it is well known that an artificial anti-correlation can be produced between dust temperature and $\beta$ because of correlated errors in the two parameters (\citealt{Shetty_2009a}; \citealt{Kelly_2012}; \citealt{Juvela_2012a}). Fig.~\ref{fig:beta_t_correlation} shows that there is a negative correlation between the measured values of $\beta$ and dust temperature for both sub-samples when assuming that the dust is optically thin. The strength of these correlations were tested with the Pearson correlation coefficient, $r_{\textrm{Pearson}}$, and Spearman's rank correlation coefficient, $\rho_{\textrm{Spearman}}$. The values ($r_{\textrm{Pearson}} = -0.83$ and $\rho_{\textrm{Spearman}} = -0.87$ for SPT; $r_{\textrm{Pearson}} = -0.89$ and $\rho_{\textrm{Spearman}} = -0.89$ for HerBS) show that the SPT and HerBS galaxies exhibit a strong negative $\beta$--$T_{\textrm{dust}}$ correlation and the similarity between the two metrics suggests that few galaxies deviate from the trend. In the following section, we address the extent to which the observed $\beta$--$T_{\textrm{dust}}$ anti-correlation is a true relationship between dust properties, reflecting an intrinsic change in the emissivity properties of dust grains with temperature, and how much is due to the fitting method. 

\begin{figure}
	\centering
	\includegraphics[width=\columnwidth]{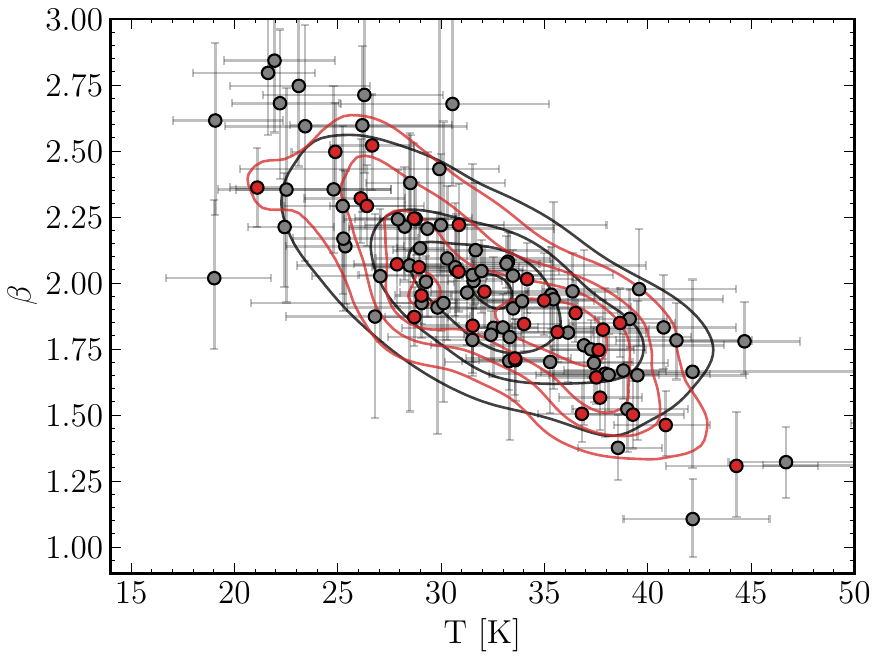}
	\caption{Relationship between the dust temperature and the emissivity index for the SPT (black) and HerBS (red) galaxies in the optically thin case.}
	\label{fig:beta_t_correlation}
\end{figure}

\subsection{Simulations}
\label{sec:simulations}

In order to assess how accurately our fitting routine derives a galaxy's dust parameters, we ran a suite of mock SEDs with known input parameters and measured how precisely we recover the dust properties from our fits. We generated these mock SEDs in the following way. We assumed that the dust emission is described by the optically-thin model (Eqn~\ref{eq:modified_blackbody_optically_thin}). A catalogue of models were produced with dust parameters randomly drawn from a uniform distribution between the lower and upper bounds in Table~\ref{tab:simulation_inputs}, chosen to reflect the width of the posterior distributions for the parameters derived for the real sources. We created two samples of mock galaxies with similar properties to the real sub-samples. To create the first sample, we placed each galaxy at a random redshift between 2 and 4 and gave the galaxy the same photometric coverage as for the HerBS sub-sample, omitting photometry at 1.2\,mm as only a few of the HerBS sources have photometry at this wavelength. To create the second sample, we placed each galaxy at a random redshift between 2 and 6 and gave the galaxy the same photometric coverage as the SPT sample. We created observed SEDs of each mock galaxy by adding flux errors to the model SED, drawing randomly from a normal distribution with a standard deviation equal to the real observational error at that wavelength. We assumed a lensing magnification of $5.5\times$ for all SPT galaxies and $5.3\times$ for all HerBS galaxies, equal to the median lensing magnification for the two sub-samples. Once we had calculated an observed SED for each mock galaxy, we removed it from the sample if it fell below the detection limits of the real samples (Section~\ref{sec:catalogue_creation}). For the sources that remained, we then derived the dust parameters using the same modelling procedure we had applied to the real sample.

\begin{table}
    \caption{Upper and lower bounds assumed on the flat priors during the simulations described in Section~\ref{sec:simulations}.}
    \centering
    \begin{tabular}{p{3cm}p{3cm}}
        \hline
        \hline
        Parameter & Bounds \\
            \hline
        log($\mu L_{\textrm{IR}} [$L$_{\odot}]$) & 13--14 (HerBS) \\
        & 13--14.5 (SPT) \\
		$T_{\textrm{dust}}$ [K] & 20--50 \\
		$\beta$  & 0.5--4 \\
		$\alpha$  & 1--5 \\
        \hline
    \end{tabular}
    \label{tab:simulation_inputs}
\end{table}

Fig.~\ref{fig:in_out_simulations} shows the paramaters derived for the two artificial samples plotted against the input values of these parameters. There is good agreement between the input and output parameters. The r.m.s.\ error (RMSE) is calculated for each dust parameter, showing the intrinsic scatter around the `true' input values. The RMSE indicate that our fitting recovers the input dust masses to 0.05\,dex and 0.03\,dex, the dust temperatures to 1.63\,K and 1.61\,K and $\beta$ to 0.13 and 0.09 of the input values for SPT and HerBS galaxies, respectively. The accuracy of the output $T_{\textrm{dust}}$ decreases for higher temperatures as a result of the SED shifting to shorter wavelengths, meaning the peak is then less constrained by the \textit{Herschel} photometry than at lower temperatures.

The right-most panel of the two simulations in Fig.~\ref{fig:in_out_simulations} shows the difference between the input and output in the $T_{\textrm{dust}}$--$\beta$ plane. Since there is no intrinsic correlation between $\beta$ and dust temperature in the model, the anti-correlation in this panel must have been introduced by correlated errors in the two parameters. The r.m.s.\ variations in $\beta$ and $T_{\textrm{dust}}$ in the artificial dataset are 0.1 and 1.6\,K, respectively. The r.m.s.\ variations in $\beta$ and $T_{\textrm{dust}}$ in the real data (Fig.~\ref{fig:beta_t_correlation}) are 0.5 and 8\,K, respectively, for the SPT galaxies, and 0.3 and 5\,K, respectively, for the HerBS galaxies. The large differences between the r.m.s.\ values for the simulation and for the real samples imply there is a genuine inverse correlation between $\beta$ and $T_{\textrm{dust}}$ which is not caused by correlated measurement errors.

\begin{figure*}
	\centering
    \includegraphics[width=\textwidth]{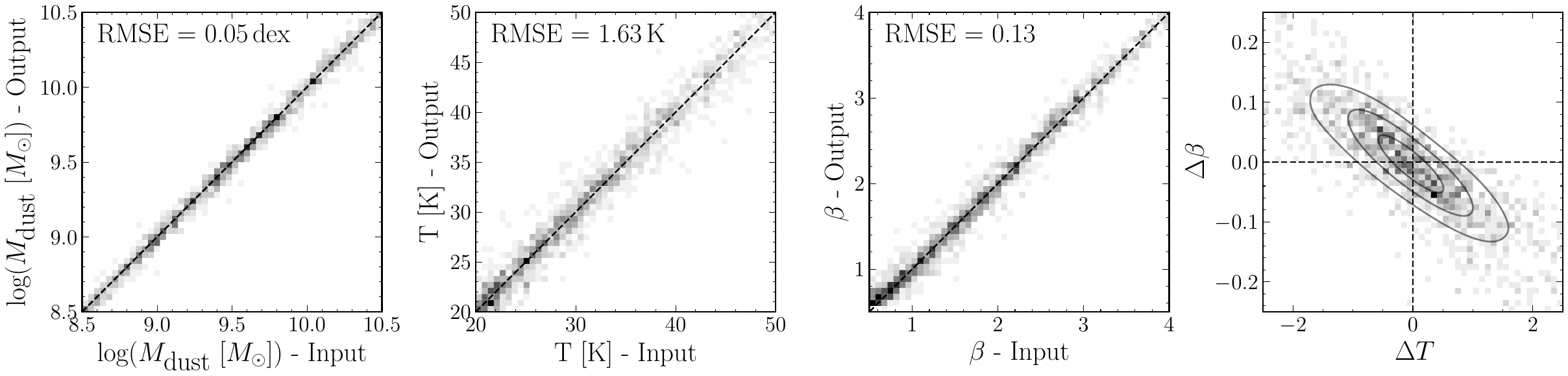}
	\includegraphics[width=\textwidth]{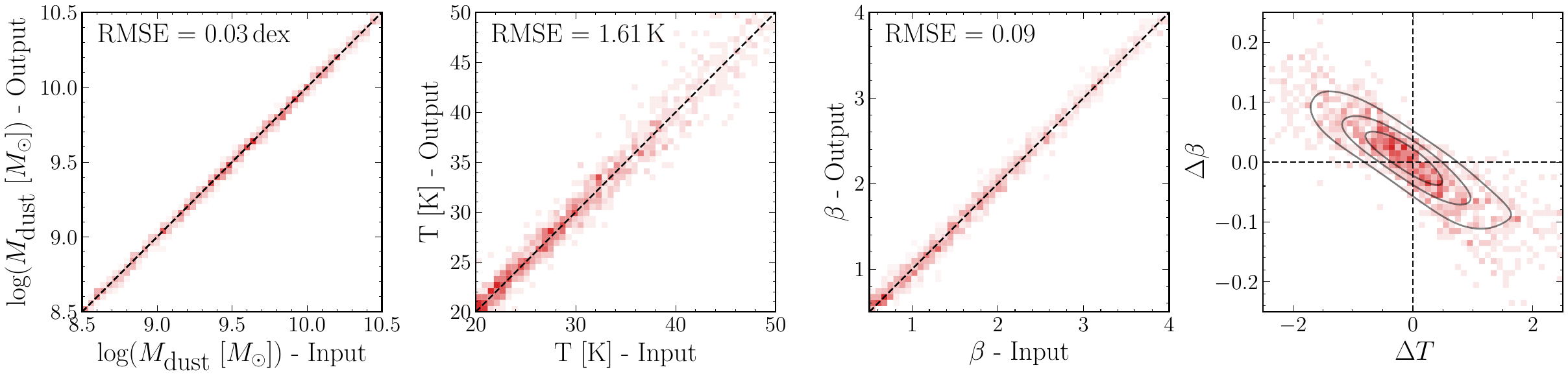}
	\caption{Comparison of the input and output (measured) values of dust mass, dust temperature and dust emissivity index for the simulations described in Section~\ref{sec:simulations}. The right-most panel shows the difference between the input and output (measured) values of dust temperature against the difference between the input and output (measured) values of $\beta$. The top row (black) shows the results for the mock SPT galaxies and the bottom row (red) shows the results for the mock HerBS galaxies.}
	\label{fig:in_out_simulations}
\end{figure*}

\section{Discussion}
\label{sec:discussion}

\subsection{Evolution of $\beta$ with redshift}
\label{sec:beta_z_evolution}

Fig.~\ref{fig:beta_z_evolution} shows the distribution of galaxies in the $\beta$--$z$ plane. The average value of the emissivity spectral index of SPT and HerBS galaxies ($\beta = 1.98$ and $1.91$) are slightly higher than the values typically assumed for galaxies in the local Universe. It is significantly higher than the value in our Galaxy, which is remarkably uniform over the sky: $\beta = 1.51\pm0.01$ (\citealt{Planck_2015}). It is also significantly higher than the average value of $\beta$ for a large sample of nearby galaxies (\citealt{Lamperti_2019}), although it is close to the values measured for the most massive galaxies in this sample, which may be the descendants of the DSFGs (\citealt{Eales_2023}). Considering the samples themselves, there is little evidence of any evolution in $\beta$ over the redshift range $2 < z < 6$. The apparent negative redshift evolution within the HerBS sample is mostly the effect of the five sources at $z \sim 4$. At this redshift, the selection wavelength of HerBS, 500\,\micron, is close to the peak of the MBB, thus biasing us towards higher temperatures. The intrinsic anti-correlation between dust temperature and $\beta$ (Section~\ref{sec:beta_t_correlation}) would then explain the low values of $\beta$ for these galaxies.

Similar conclusions have been reached in two other recent studies. For an independent sample of DSFGs selected from HerBS, \citet{Ismail_2023} found no evolution in $\beta$ over $2 < z < 3$. For a sample of 17 galaxies, \citet{Witstok_2023} found no evolution across $4 < z < 8$.

\begin{figure*}
	\centering
	\includegraphics[width=\textwidth]{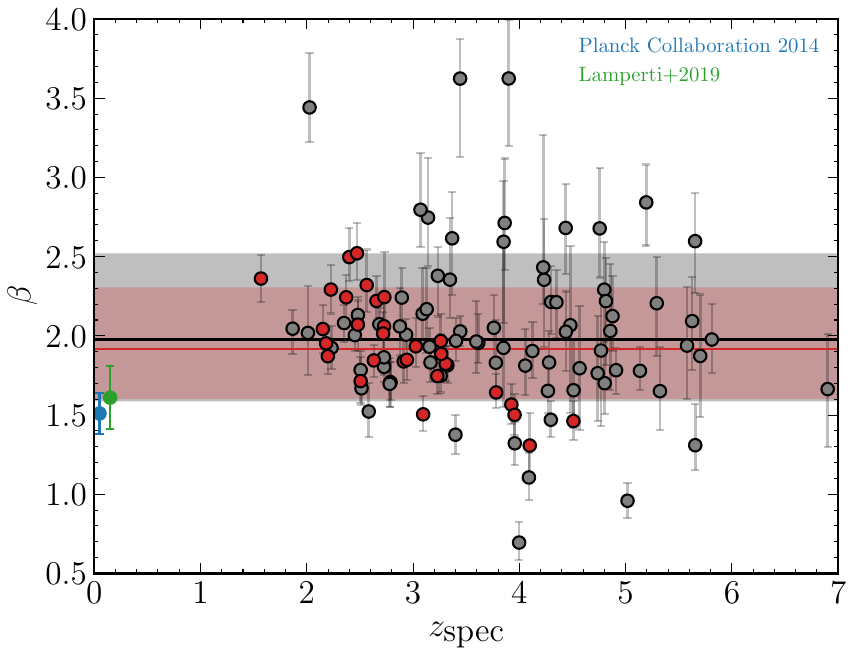}
	\caption{Distribution of $\beta$ for the HerBS (red) and SPT (black) galaxies versus redshift, with the canonical values from JINGLES (green) and the Milky Way (blue). The median value of the stacked posterior distributions are shown as red and black lines for the HerBS and SPT sub-samples, respectively, with the shaded regions showing the 16th to 84th percentiles.}
	\label{fig:beta_z_evolution}
\end{figure*}

\subsection{Variation in $\beta$}

To assess whether the variation of $\beta$ in Fig.~\ref{fig:beta_z_evolution} is due to genuine differences in the properties of dust between DSFGs, or whether it is simply due to measurement errors, we repeated the simulations of the previous section. This time we used values of $\beta$ drawn from a uniform distribution covering the $\pm1\sigma$ range of the observed values in Fig.~\ref{fig:beta_z_evolution}, generating -- as before -- two samples of mock galaxies with observational data similar to that of the SPT and HerBS sub-samples.

Fig.~\ref{fig:beta_z_simulation} shows a comparison of the input and output values of $\beta$ as a function of redshift for both samples. There is no tendency for our fitting method to over- or under-predict the true value of $\beta$ at any redshift. The r.m.s.\ scatter in Fig.~\ref{fig:beta_z_simulation} is $\simeq$0.1, which is much less than the r.m.s.\ scatter in $\beta$ in Fig.~\ref{fig:beta_z_evolution} ($\simeq$0.3--0.5), which shows that the measurement errors are much smaller than the scatter in Fig.~\ref{fig:beta_z_evolution}. Fig.~\ref{fig:beta_z_evolution} therefore shows that there are real differences in the properties of dust between DSFGs, a phenomenon also seen in the low-redshift galaxy population (\citealt{Lamperti_2019}).

\begin{figure}
	\centering
	\includegraphics[width=\columnwidth]{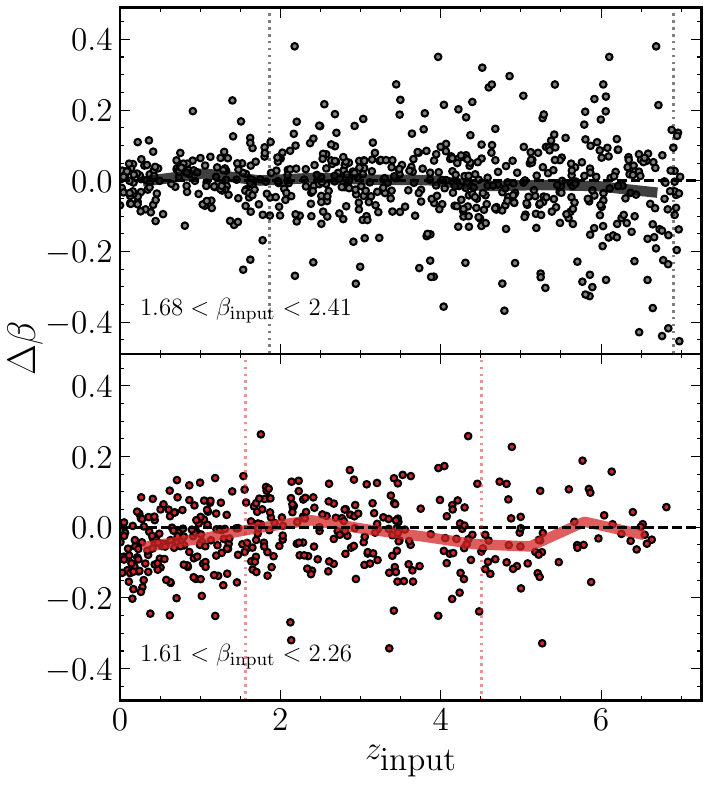}
	\caption{Difference between the input and output $\beta$ values, $\Delta \beta = \beta_{\textrm{output}} - \beta_{\textrm{input}}$, as a function of the input redshift for mock SPT galaxies (top panel) and mock \textit{Herschel} galaxies (bottom panel). Filled circles represent mock galaxies with SEDs that would have been observed according to their respective survey limits. Dotted vertical lines represent the minimum and maximum redshift observed in each sub-sample. The thick solid lines represent the median in $\Delta \beta$.}
	\label{fig:beta_z_simulation}
\end{figure}

In their study of an independent sample of DSFGs drawn from HerBS, \citet{Ismail_2023} reached similar conclusions. Across $2 < z <4$, they found an average value of $\beta = 2.2\pm0.2$ with a range of $\beta$ from 1.5 to 3.0, very similar to our results for the combined HerBS/SPT sample. They, too, concluded that the variation in $\beta$ is caused by intrinsic differences between the dust in different DSFGs rather than by measurement errors. In their sample of 17 galaxies in the range $4 < z < 8$, \citet{Witstok_2023} found a mean value of $\beta = 1.8\pm0.3$, in agreement with our estimates, with a similar wide range of $\beta$ from $\simeq1.5$ to $\simeq2.7$. Other recent studies of high-redshift galaxies find similar values of $\beta$ (\citealt{daCunha_2021, Cooper_2022}).

Some of the SPT galaxies in Fig.~\ref{fig:beta_z_evolution} have values of $\beta \simeq3$, as do galaxies in other high-redshift samples (\citealt{Ismail_2023, Witstok_2023}). Such values are higher than the global values seen in local galaxies (\citealt{Lamperti_2019}), although such high values have been seen within Andromeda (\citealt{Smith_2012}) and the Galaxy (\citealt{Bracco_2011}). Are there any methodological limitations which might explain these results?

For our sample, one possible explanation might be the 3-mm photometry for the SPT sources. These flux densities were measured from ALMA images, so it is possible that they are too low because of the possibility -- as with all interferometers -- that some extended flux has been missed (the `zero baseline problem'). We assessed how much would need to be lost by repeating the SED fitting without the 3-mm photometry, with $\beta = 2$. We did this for five sources: SPT0112$-$55 ($\beta = 3.62$), SPT0611$-$55 ($\beta = 3.44$), SPT2203$-$41 ($\beta = 2.84$), SPT2349$-$52 ($\beta = 3.62$) and SPT2357$-$51 ($\beta = 2.80$). Comparing the flux density at 3\,mm predicted by the model with the observed flux density, we find that the observed flux density would need to be between 60--80 per cent lower than the true flux density to explain the very high values of $\beta$. Although this is possible, given the structures seen for high-redshift galaxies it seems unlikely, and it seems more likely to us that some of these extreme values found by us and others are genuine.

\subsection{The inverse $\beta$--$T_{\textrm{dust}}$ relationship}

Our simulations imply that the inverse relationship seen between $\beta$ and $T_{\textrm{dust}}$ is not the result of measurement errors. Another study of an independent sample of HerBS sources reached a similar conclusion (\citealt{Ismail_2023}). A similar conclusion was also reached by a hierarchical Bayesian study of nearby galaxies (\citealt{Lamperti_2019}).

\subsection{Evolution of $T_{\textrm{dust}}$ with redshift}

Does the temperature of dust increase with redshift? This has been a matter of intense debate in the literature. Some groups find evidence for an increase, e.g. \citet{Magdis_2012,Magnelli_2014,Swinbank_2014,Bethermin_2015,Ivison_2016,Faisst_2017,Schreiber_2018,Zavala_2018b,Liang_2019,Ma_2019,Faisst_2020,Bakx_2021,Witstok_2023}, which might not be surprising since both high specific star-formation rates and lower dust abundances at high redshift might lead to higher dust temperatures. Others report a fall, e.g.~\citet{Symeonidis_2013}. An important caveat, noted by many of those studies, is the correlation between dust temperature and luminosity (\citealt{Dunne_2000}) which -- since galaxies at high redshift tend to have high luminosities -- might lead to a correlation between dust temperature and redshift. In contrast, the works of \citet{Casey_2018,Jin_2019,Lim_2020,Dudzeviciute_2020,Reuter_2020,Barger_2022,Drew_2022} and \citet{Witstok_2023}, among others, report little or no evolution of temperature with redshift at a given fixed IR luminosity. 

The dust temperature derived from SED fitting depends on the model assumed for the dust opacity, whereas $\lambda_{\textrm{peak}}$ is the same for every model. To avoid this dependence, we have derived $\lambda_{\textrm{peak}}$ for each source and then calculated $T_{\textrm{dust}}$ from Wien's displacement law. Although this will not be the same as the true temperature of the dust, it allows us to see whether there is any evidence for a change in $T_{\textrm{dust}}$ with redshift.

The top panel of Fig.~\ref{fig:t_evolution} shows the distribution of IR luminosity for the full sample of galaxies as obtained from the fitting of Equation~4, as well as the detection limits. In calculating the detection limits, we have assumed an SED with a dust temperature of 32\,K, the median temperature of the sample, and $\beta = 2$. To avoid the selection effect caused by the luminosity--temperature relationship, we only consider sources selected in a small range of IR luminosity, as illustrated by the boxed regions. The bottom panel of Fig.~\ref{fig:t_evolution} shows the peak dust temperature ($T_{\textrm{peak}} = 2.898 \times 10^{3}$ [\micron K]/$\lambda_{\textrm{peak}}$ [\micron]) versus redshift. We have plotted the observational relationships inferred by \citet{Schreiber_2018}, \citet{Bouwens_2020} and \citet{Viero_2022}. Bootstrap fitting with a linear model gives an evolution rate of $5.20\pm0.92$ and $-0.25\pm0.52$ $T_{\textrm{peak}}/z$ for the HerBS and SPT galaxies, respectively. In a study of an independent sample of HerBS galaxies, \citet{Ismail_2023} found a rate of evolution of $6.5\pm0.5\,K/z$, similar to our value.

 While there is statistical evidence within the HerBS sub-sample -- both ours and that of \citet{Ismail_2023} -- for the evolution of temperature with redshift, there seems little evidence for any such evolution over the redshift range covered by the combined HerBS and SPT sample.
 
 We suggest there is a simple selection effect that explains the temperature evolution seen in the HerBS sample. For the highest redshift sources in the HerBS sub-sample, the selection wavelength in the rest frame is close to the peak of the MBB, which leads to a selection bias towards sources with higher temperatures. Given the lack of any obvious correlation when both sub-samples are considered together, we conclude that there is no strong evidence for temperature evolution. A plausible suggestion for the disagreement in the literature was suggested by \citet{Drew_2022}: the studies that have found no evolution are those that looked for evolution at a fixed infrared luminosity, as in this paper, whereas those that found evolution were generally looking for evolution at a fixed stellar mass. The latter would be expected to see evolution because of the gradual increase in specific star-formation rate -- and thus the increase in the intensity of the interstellar radiation field -- with redshift.

\begin{figure}
	\centering
	\includegraphics[width=0.97\columnwidth]{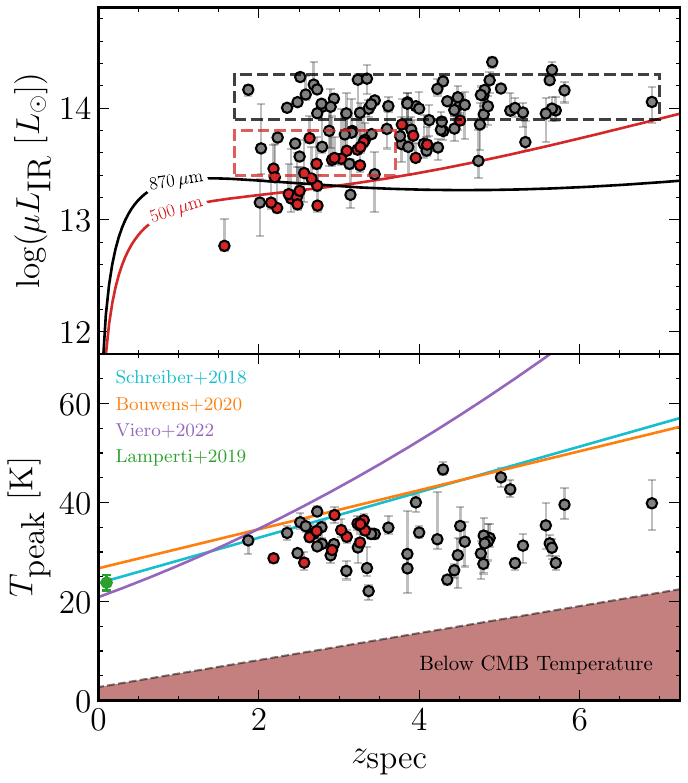}
	\caption{Top: distribution of HerBS (red) and SPT (black) galaxies in the $\mu L_{\textrm{IR}}$--$z$ plane. Red and black lines represent the detection limit of a source detected at $> 80$\,mJy (500\,\micron) and $> 25$\,mJy (870\,\micron) with $\beta = 2$ and $T_{\textrm{dust}} = 32$\,K. Boxed regions show the limits of the unbiased, luminosity-limited set of sources from which we test evolutionary trends in the dust temperature. Bottom: distribution of dust temperature with redshift for the sources selected in our luminosity-limited subset. Observed trends from \citet{Schreiber_2018,Bouwens_2020} and \citet{Viero_2022} are illustrated as cyan, orange and purple lines, respectively. The median dust temperature of local sources from JINGLES is shown in green.}
	\label{fig:t_evolution}
\end{figure}

\section{Conclusions}
\label{sec:conclusions}

In this paper we estimate values of $\beta$ and dust temperature for a sample of 109 dusty star-forming galaxies spanning $2 < z < 6$.  We have obtained the following results:

\begin{itemize}

\item We have compared the results from fitting the emission from the galaxies with an optically-thin model and with two general-opacity models, one in which the dust becomes optically thick at a wavelength of 100\,\micron, the other where the dust becomes optically thick at 200\,\micron. We find that our estimates of $\beta$, the peak of the galaxy SED, and the far-infrared luminosity, are similar between the models, although there is a change in $\beta$ of 0.2 between the most extreme models. In contrast, our estimates of dust mass and dust temperature depend strongly on the opacity assumptions.

\item For a sub-sample of 37 galaxies, we have measurements of the sizes of the sources, which allows us to estimate the opacity directly. We find very similar estimates of the galaxy parameters with this model to the optically-thin model, which we use for the rest of the paper.

\item We find no evidence of any change in $\beta$ with redshift across $2 < z < 6$, with a median value of $\beta = 1.96$ for the optically-thin model and a 14--86\% confidence interval for the population of 1.65 to 2.38. Using simulations, we estimate the measurement errors from our procedure and show that the variation of $\beta$ in the population is the result of intrinsic variations in the properties of the dust in DSFGs, the same found between galaxies in the low-redshift Universe (\citealt{Lamperti_2019}). 

\item After allowing for the effect of correlated errors, we find an inverse correlation between our estimates of $\beta$ and dust temperature.

\item We find no evidence at a fixed far-infrared luminosity for any change in dust temperature with redshift.

\end{itemize}

The results in this and other recent papers show that the properties of dust appear to differ between galaxies, both at high and low redshift. Ours and other recent papers show that there appears to be intrinsic anti-correlation between dust emissivity index and dust temperature. As far as we know, there is no chemical/physical model of dust that explains either of these results.

\section*{Acknowledgements}

BW thanks the Engineering and Physical Sciences Research Council for a PhD studentship. SAE thanks the Science and Technology Facilities Council (consolidated grant ST/K000926/1) for the funds that supported this research.

%%%%%%%%%%%%%%%%%%%%%%%%%%%%%%%%%%%%%%%%%%%%%%%%%%

\section*{Data availability}

The photometry for each galaxy and the results of fitting our models to the
photometry are listed in the appendix.



\bibliographystyle{mnras}
\bibliography{bibliography} % if your bibtex file is called example.bib

%\begin{thebibliography}{99}
%\end{thebibliography}

%%%%%%%%%%%%%%%%%%%%%%%%%%%%%%%%%%%%%%%%%%%%%%%%%%

%%%%%%%%%%%%%%%%%%%%%%%%%%%%%%%%%%%%%%%%%%%%%%%%%%

% Don't change these lines
\bsp	% typesetting comment

%%%%%%%%%%%%%%%%%%%%%%%%%%%%%%%%%%%%%%%%%%%%%%%%%%

%%%%%%%%%%%%%%%%% APPENDICES %%%%%%%%%%%%%%%%%%%%%

\appendix

\onecolumn
\section{HerBS photometry}

\begin{longtable}{lccccccccr}

	\caption{Spectroscopic redshifts, lensing magnifications and photometry for HerBS sources.}
	\label{tab:data_herbs} \\
	\hline
	\hline
	Source & Spec-z & $\mu$ & $S_{250\,\micron}$ & $S_{350\,\micron}$ & $S_{500\,\micron}$ & $S_{850\,\micron}$ & $S_{1.2\,\textrm{mm}}$ & $S_{2\,\textrm{mm}}$ & $S_{3\,\textrm{mm}}$ \\
    & & & (mJy) & (mJy) & (mJy) & (mJy) & (mJy) & (mJy) & (mJy) \\
	\hline
	HerBS-11 & 2.631 & 18.4 & 257.5$\pm$6.4 & 271.1$\pm$6.3 & 204.0$\pm$7.2 & 67.3$\pm$6.3 & -- & 3.59$\pm$0.03 & 0.93$\pm$0.02\\
	HerBS-14 & 3.782 & 36.4 & 116.3$\pm$6.1 & 177.0$\pm$6.3 & 179.3$\pm$7.5 & 77.9$\pm$6.4 & 28.9$\pm$0.6 & 7.31$\pm$0.04 & 1.50$\pm$0.02\\
	HerBS-18 & 2.182 & 27.9 & 212.9$\pm$4.7 & 244.2$\pm$5.0 & 169.4$\pm$6.2 & 52.9$\pm$6.1 & -- & 2.73$\pm$0.03 & 0.74$\pm$0.06\\
	HerBS-21 & 3.323 & 6.1 & 125.8$\pm$5.5 & 185.5$\pm$5.8 & 155.1$\pm$7.4 & 51.3$\pm$6.3 & -- & 3.92$\pm$0.03 & 0.94$\pm$0.10\\
	HerBS-24 & 2.198 & 4.6 & 170.9$\pm$5.7 & 197.1$\pm$6.3 & 145.6$\pm$7.4 & 64.8$\pm$7.8 & -- & 2.68$\pm$0.03 & 0.74$\pm$0.03\\
	HerBS-25 & 2.912 & 49.8 & 112.5$\pm$5.0 & 148.0$\pm$5.4 & 143.4$\pm$6.5 & 49.2$\pm$5.7 & -- & 3.52$\pm$0.03 & 0.81$\pm$0.11\\
	HerBS-27 & 4.509 & 14.6 & 72.2$\pm$5.3 & 129.8$\pm$5.6 & 138.6$\pm$7.0 & 90.5$\pm$6.3 & 28.9$\pm$1.0 & 8.70$\pm$0.03 & 1.97$\pm$0.02\\
	HerBS-28 & 3.925 & 5.7 & 79.4$\pm$5.8 & 135.4$\pm$6.0 & 140.0$\pm$7.4 & 79.4$\pm$7.8 & 25.5$\pm$1.2 & 5.56$\pm$0.03 & 1.66$\pm$0.10\\
	HerBS-36 & 3.095 & 5.4 & 121.5$\pm$6.1 & 161.0$\pm$6.7 & 125.5$\pm$7.7 & 64.0$\pm$8.8 & -- & 4.81$\pm$0.03 & 1.11$\pm$0.02\\
	HerBS-39 & 3.229 & 5.1 & 118.3$\pm$5.1 & 141.2$\pm$5.5 & 119.7$\pm$6.8 & 36.5$\pm$7.0 & -- & 3.01$\pm$0.03 & 0.64$\pm$0.03\\
	HerBS-41 & 4.098 & 2.1 & 63.3$\pm$6.2 & 91.1$\pm$6.1 & 121.7$\pm$7.4 & 31.8$\pm$5.9 & 15.2$\pm$0.5 & 4.85$\pm$0.07 & --\\
	HerBS-42 & 3.307 & 3.9 & 130.3$\pm$5.8 & 160.0$\pm$6.1 & 116.2$\pm$6.8 & 50.4$\pm$6.1 & -- & 2.95$\pm$0.06 & 0.58$\pm$0.05\\
	HerBS-49 & 2.727 & 15.3 & 76.8$\pm$6.0 & 110.9$\pm$6.2 & 110.4$\pm$7.3 & 31.9$\pm$8.5 & -- & 1.69$\pm$0.04 & 1.13$\pm$0.09\\
	HerBS-55 & 2.656 & 13.7 & 109.0$\pm$5.3 & 116.5$\pm$5.5 & 107.1$\pm$6.6 & 29.9$\pm$7.1 & -- & 1.05$\pm$0.03 & 0.33$\pm$0.02\\
	HerBS-57 & 3.265 & 15.0 & 118.1$\pm$4.9 & 147.3$\pm$5.2 & 105.4$\pm$6.4 & 60.7$\pm$8.5 & -- & 3.05$\pm$0.02 & 0.50$\pm$0.02\\
	HerBS-60 & 3.261 & 9.5 & 73.3$\pm$5.8 & 101.2$\pm$6.1 & 103.6$\pm$7.5 & 39.8$\pm$5.7 & 13.2$\pm$0.8 & 2.54$\pm$0.03 & 0.53$\pm$0.02\\
	HerBS-68 & 2.719 & 10.5 & 139.1$\pm$5.3 & 144.8$\pm$5.4 & 100.5$\pm$6.6 & 46.7$\pm$7.7 & -- & 1.62$\pm$0.04 & 0.32$\pm$0.05\\
	HerBS-73 & 3.026 & 3.1 & 117.1$\pm$6.0 & 129.0$\pm$6.2 & 99.6$\pm$7.4 & 49.4$\pm$6.8 & -- & 2.09$\pm$0.03 & 0.43$\pm$0.02\\
	HerBS-86 & 2.564 & 4.8 & 77.4$\pm$5.6 & 90.7$\pm$5.8 & 96.0$\pm$7.4 & 33.8$\pm$6.1 & 8.1$\pm$0.6 & 1.56$\pm$0.02 & 0.25$\pm$0.02\\
	HerBS-93 & 2.400 & 2.4 & 77.3$\pm$5.4 & 87.3$\pm$5.7 & 94.8$\pm$7.0 & 22.8$\pm$5.7 & 6.2$\pm$0.4 & 1.39$\pm$0.02 & 0.16$\pm$0.01\\
	HerBS-103 & 2.942 & 4.3 & 126.1$\pm$5.3 & 131.2$\pm$5.7 & 93.5$\pm$7.0 & 33.7$\pm$7.6 & -- & 1.40$\pm$0.03 & 0.42$\pm$0.02\\
	HerBS-111 & 2.371 & 4.3 & 105.9$\pm$6.5 & 115.6$\pm$6.2 & 92.7$\pm$7.4 & 13.3$\pm$6.9 & -- & 1.13$\pm$0.02 & 0.21$\pm$0.02\\
	HerBS-132 & 2.473 & 4.8 & 86.7$\pm$5.8 & 102.6$\pm$6.0 & 90.6$\pm$7.8 & 16.6$\pm$7.4 & -- & 0.86$\pm$0.02 & 0.16$\pm$0.02\\
	HerBS-145 & 2.730 & 1.9 & 54.7$\pm$6.0 & 67.4$\pm$6.2 & 86.8$\pm$7.7 & 13.4$\pm$7.0 & 5.1$\pm$0.6 & 1.06$\pm$0.04 & --\\
	HerBS-160 & 3.955 & 8.1 & 48.6$\pm$5.6 & 84.2$\pm$6.0 & 84.8$\pm$7.1 & 36.6$\pm$6.4 & 15.0$\pm$0.6 & 3.94$\pm$0.02 & 0.91$\pm$0.02\\
	HerBS-182 & 2.227 & 1.1 & 89.0$\pm$5.7 & 109.1$\pm$6.2 & 82.3$\pm$7.9 & -- & -- & 1.04$\pm$0.03 & 0.21$\pm$0.02\\
	HerBS-184 & 2.507 & 6.0 & 91.9$\pm$5.9 & 107.6$\pm$6.0 & 82.3$\pm$7.1 & -- & -- & 1.49$\pm$0.02 & 0.48$\pm$0.02\\
	HerBS-200 & 2.151 & 0.4 & 107.1$\pm$6.1 & 109.7$\pm$6.0 & 80.5$\pm$7.5 & 11.3$\pm$6.6 & -- & 0.78$\pm$0.02 & 0.25$\pm$0.01\\
	HerBS-207 & 1.569 & 5.2 & 96.9$\pm$5.9 & 121.7$\pm$6.1 & 80.2$\pm$7.5 & 33.1$\pm$6.3 & -- & 0.89$\pm$0.02 & 0.25$\pm$0.03\\
	HerBS-208 & 2.481 & 1.4 & 69.4$\pm$5.1 & 91.9$\pm$5.5 & 80.1$\pm$6.6 & -- & -- & 1.40$\pm$0.05 & 0.32$\pm$0.04\\
	\hline
\end{longtable}

\newpage

\section{Best-fitting SED parameters}

\begin{longtable}{lccccr}
	\caption{Best-fitting SED parameters for all SPT sources.}
	\label{tab:parameters_spt} \\
	\hline
	\hline
    Source & Spec-z & log($M_{\textrm{dust}}$ [$M_\odot$]) & $T_{\textrm{dust}}$ [K] & $\beta$ & $\chi^2_\nu$ \\
	\hline
	\endhead
	SPT0002-52 & 2.351 & $8.95\substack{+0.05\\-0.05}$ & $33.23\substack{+2.19\\-1.80}$ & $2.08\substack{+0.11\\-0.12}$ & 2.05\\
	SPT0020-51 & 4.123 & $9.16\substack{+0.06\\-0.06}$ & $33.49\substack{+3.02\\-3.04}$ & $1.90\substack{+0.18\\-0.17}$ & 0.57\\
	SPT0027-50 & 3.444 & $9.29\substack{+0.04\\-0.04}$ & $33.48\substack{+1.35\\-1.25}$ & $2.03\substack{+0.10\\-0.09}$ & 3.02\\
	SPT0054-41 & 4.877 & $9.26\substack{+0.08\\-0.06}$ & $31.67\substack{+3.54\\-3.99}$ & $2.12\substack{+0.25\\-0.22}$ & 0.84\\
	SPT0103-45 & 3.090 & $9.57\substack{+0.08\\-0.07}$ & $25.37\substack{+2.84\\-2.88}$ & $2.14\substack{+0.29\\-0.22}$ & 1.10\\
	SPT0106-64 & 4.910 & $9.21\substack{+0.05\\-0.05}$ & $41.39\substack{+3.65\\-2.93}$ & $1.78\substack{+0.14\\-0.15}$ & 3.65\\
	SPT0109-47 & 3.614 & $8.78\substack{+0.05\\-0.05}$ & $35.33\substack{+2.12\\-3.32}$ & $1.96\substack{+0.18\\-0.13}$ & 2.85\\
	SPT0112-55 & 3.443 & $9.18\substack{+0.24\\-0.35}$ & $14.35\substack{+4.40\\-1.59}$ & $3.62\substack{+0.25\\-0.50}$ & 6.53\\
	SPT0113-46 & 4.233 & $8.84\substack{+0.13\\-0.12}$ & $22.52\substack{+5.08\\-3.33}$ & $2.35\substack{+0.38\\-0.43}$ & 1.40\\
	SPT0125-47 & 2.515 & $9.47\substack{+0.05\\-0.05}$ & $38.81\substack{+2.07\\-2.88}$ & $1.67\substack{+0.14\\-0.11}$ & 2.69\\
	SPT0125-50 & 3.957 & $8.77\substack{+0.06\\-0.06}$ & $46.69\substack{+3.35\\-2.78}$ & $1.32\substack{+0.13\\-0.13}$ & 3.83\\
	SPT0136-63 & 4.299 & $9.18\substack{+0.07\\-0.06}$ & $28.24\substack{+2.68\\-2.68}$ & $2.21\substack{+0.22\\-0.20}$ & 0.42\\
	SPT0147-64 & 4.803 & $9.30\substack{+0.07\\-0.06}$ & $35.28\substack{+3.98\\-5.08}$ & $1.70\substack{+0.27\\-0.19}$ & 2.39\\
	SPT0150-59 & 2.788 & $9.26\substack{+0.05\\-0.05}$ & $33.29\substack{+1.88\\-2.26}$ & $1.70\substack{+0.13\\-0.11}$ & 2.63\\
	SPT0155-62 & 4.349 & $10.01\substack{+0.07\\-0.07}$ & $22.43\substack{+2.40\\-1.78}$ & $2.21\substack{+0.20\\-0.23}$ & 3.32\\
	SPT0202-61 & 5.018 & $9.12\substack{+0.04\\-0.04}$ & $57.61\substack{+4.19\\-3.64}$ & $0.96\substack{+0.11\\-0.11}$ & 6.46\\
	SPT0226-45 & 3.233 & $9.01\substack{+0.07\\-0.08}$ & $28.51\substack{+4.58\\-2.15}$ & $2.38\substack{+0.18\\-0.26}$ & 4.29\\
	SPT0243-49 & 5.702 & $9.67\substack{+0.11\\-0.09}$ & $26.80\substack{+4.61\\-4.30}$ & $1.87\substack{+0.39\\-0.38}$ & 0.71\\
	SPT0245-63 & 5.626 & $9.84\substack{+0.08\\-0.08}$ & $30.30\substack{+4.99\\-3.47}$ & $2.09\substack{+0.28\\-0.31}$ & 1.07\\
	SPT0300-46 & 3.595 & $9.35\substack{+0.07\\-0.07}$ & $31.26\substack{+2.94\\-3.23}$ & $1.96\substack{+0.26\\-0.20}$ & 0.96\\
	SPT0311-58 & 6.901 & $9.28\substack{+0.09\\-0.08}$ & $42.17\substack{+10.20\\-7.33}$ & $1.66\substack{+0.35\\-0.37}$ & 1.37\\
	SPT0314-44 & 2.935 & $9.44\substack{+0.04\\-0.04}$ & $31.57\substack{+1.35\\-1.22}$ & $2.01\substack{+0.10\\-0.09}$ & 2.96\\
	SPT0319-47 & 4.510 & $9.38\substack{+0.09\\-0.08}$ & $37.93\substack{+5.07\\-5.60}$ & $1.66\substack{+0.33\\-0.24}$ & 1.63\\
	SPT0345-47 & 4.296 & $8.85\substack{+0.06\\-0.06}$ & $52.45\substack{+2.45\\-2.60}$ & $1.47\substack{+0.12\\-0.11}$ & 1.22\\
	SPT0346-52 & 5.655 & $9.24\substack{+0.05\\-0.06}$ & $52.76\substack{+4.10\\-7.17}$ & $1.31\substack{+0.26\\-0.16}$ & 2.94\\
	SPT0348-62 & 5.654 & $9.59\substack{+0.14\\-0.12}$ & $26.19\substack{+4.35\\-3.48}$ & $2.60\substack{+0.30\\-0.33}$ & 0.21\\
	SPT0402-45 & 2.683 & $9.46\substack{+0.04\\-0.04}$ & $33.16\substack{+1.60\\-1.62}$ & $2.07\substack{+0.10\\-0.09}$ & 8.22\\
	SPT0403-58 & 4.056 & $9.40\substack{+0.07\\-0.07}$ & $36.14\substack{+3.77\\-4.35}$ & $1.81\substack{+0.23\\-0.19}$ & 2.28\\
	SPT0418-47 & 4.225 & $8.30\substack{+0.19\\-0.09}$ & $29.92\substack{+2.89\\-9.65}$ & $2.43\substack{+0.83\\-0.23}$ & 7.44\\
	SPT0425-40 & 5.135 & $8.78\substack{+0.05\\-0.05}$ & $44.68\substack{+2.69\\-3.05}$ & $1.78\substack{+0.15\\-0.12}$ & 4.65 \\
	SPT0436-40 & 3.852 & $9.20\substack{+0.16\\-0.16}$ & $23.42\substack{+7.81\\-3.87}$ & $2.59\substack{+0.38\\-0.51}$ & 2.63\\
	SPT0441-46 & 4.480 & $8.89\substack{+0.13\\-0.13}$ & $28.48\substack{+11.44\\-5.46}$ & $2.07\substack{+0.50\\-0.55}$ & 3.66\\
	SPT0452-50 & 2.011 & $10.07\substack{+0.13\\-0.10}$ & $19.02\substack{+2.76\\-2.34}$ & $2.02\substack{+0.30\\-0.27}$ & 1.63\\
	SPT0459-58 & 4.856 & $9.06\substack{+0.11\\-0.09}$ & $31.52\substack{+6.78\\-5.52}$ & $2.03\substack{+0.42\\-0.37}$ & 0.68\\
	SPT0459-59 & 4.799 & $9.48\substack{+0.09\\-0.09}$ & $25.24\substack{+3.93\\-2.83}$ & $2.29\substack{+0.30\\-0.33}$ & 0.80\\
	SPT0512-59 & 2.233 & $9.39\substack{+0.04\\-0.05}$ & $29.06\substack{+1.69\\-1.46}$ & $1.92\substack{+0.14\\-0.13}$ & 0.83\\
	SPT0516-59 & 3.404 & $8.68\substack{+0.05\\-0.05}$ & $36.36\substack{+2.21\\-2.30}$ & $1.97\substack{+0.15\\-0.13}$ & 2.31\\
	SPT0520-53 & 3.779 & $9.09\substack{+0.08\\-0.07}$ & $32.54\substack{+3.22\\-4.43}$ & $1.83\substack{+0.27\\-0.18}$ & 1.54\\
	SPT0528-53 & 4.737 & $8.69\substack{+0.12\\-0.08}$ & $36.91\substack{+6.77\\-7.38}$ & $1.76\substack{+0.36\\-0.30}$ & 1.14\\
	SPT0529-54 & 3.368 & $9.38\substack{+0.09\\-0.11}$ & $19.07\substack{+3.29\\-2.03}$ & $2.62\substack{+0.29\\-0.36}$ & 0.80\\
	SPT0532-50 & 3.399 & $9.33\substack{+0.04\\-0.04}$ & $38.56\substack{+2.15\\-1.94}$ & $1.38\substack{+0.13\\-0.12}$ & 0.86\\
	SPT0538-50 & 2.786 & $8.98\substack{+0.05\\-0.06}$ & $33.59\substack{+1.98\\-1.75}$ & $1.71\substack{+0.15\\-0.16}$ & 1.36\\
	SPT0544-40 & 4.269 & $9.14\substack{+0.05\\-0.05}$ & $38.11\substack{+2.45\\-2.89}$ & $1.65\substack{+0.15\\-0.14}$ & 2.75\\
	SPT0550-53 & 3.128 & $9.20\substack{+0.08\\-0.09}$ & $25.27\substack{+3.60\\-2.46}$ & $2.17\substack{+0.26\\-0.28}$ & 1.92\\
	SPT0551-48 & 2.583 & $9.54\substack{+0.06\\-0.06}$ & $39.01\substack{+2.94\\-2.66}$ & $1.52\substack{+0.18\\-0.16}$ & 2.27\\
	SPT0551-50 & 3.164 & $9.24\substack{+0.05\\-0.05}$ & $32.99\substack{+1.71\\-1.58}$ & $1.83\substack{+0.13\\-0.12}$ & 2.83\\
	SPT0552-42 & 4.438 & $9.25\substack{+0.10\\-0.09}$ & $27.06\substack{+3.98\\-3.29}$ & $2.03\substack{+0.25\\-0.25}$ & 0.88\\
	SPT0553-50 & 5.323 & $8.89\substack{+0.07\\-0.07}$ & $39.50\substack{+5.27\\-4.90}$ & $1.65\substack{+0.28\\-0.25}$ & 0.90\\
	SPT0555-62 & 4.815 & $9.01\substack{+0.10\\-0.10}$ & $30.00\substack{+8.01\\-3.66}$ & $2.22\substack{+0.27\\-0.40}$ & 4.84\\
	SPT0604-64 & 2.481 & $9.52\substack{+0.04\\-0.04}$ & $28.98\substack{+1.14\\-1.11}$ & $2.13\substack{+0.09\\-0.08}$ & 4.53\\
	SPT0611-55 & 2.026 & $8.92\substack{+0.07\\-0.07}$ & $19.45\substack{+1.33\\-2.26}$ & $3.44\substack{+0.34\\-0.22}$ & 4.45\\
	SPT0625-58 & 2.727 & $9.48\substack{+0.04\\-0.04}$ & $32.41\substack{+1.44\\-1.32}$ & $1.80\substack{+0.09\\-0.09}$ & 2.85\\
	SPT0652-55 & 3.347 & $9.65\substack{+0.14\\-0.08}$ & $24.80\substack{+2.77\\-4.72}$ & $2.35\substack{+0.39\\-0.24}$ & 4.53\\
	SPT2031-51 & 2.452 & $9.40\substack{+0.04\\-0.04}$ & $29.27\substack{+1.37\\-1.26}$ & $2.00\substack{+0.10\\-0.10}$ & 2.18\\
	SPT2037-65 & 3.998 & $9.89\substack{+0.04\\-0.04}$ & $46.72\substack{+3.06\\-3.15}$ & $0.70\substack{+0.13\\-0.11}$ & 2.16\\
	SPT2048-55 & 4.090 & $9.24\substack{+0.05\\-0.06}$ & $42.17\substack{+3.71\\-3.35}$ & $1.11\substack{+0.15\\-0.14}$ & 1.07\\
	SPT2101-60 & 3.155 & $9.11\substack{+0.04\\-0.04}$ & $33.91\substack{+1.73\\-1.70}$ & $1.93\substack{+0.12\\-0.11}$ & 1.76\\
	SPT2103-60 & 4.436 & $8.65\substack{+0.10\\-0.09}$ & $22.20\substack{+2.88\\-2.33}$ & $2.68\substack{+0.28\\-0.29}$ & 2.44\\
	SPT2129-57 & 3.260 & $9.04\substack{+0.05\\-0.05}$ & $37.26\substack{+2.08\\-1.88}$ & $1.75\substack{+0.11\\-0.11}$ & 3.51\\
	SPT2132-58 & 4.768 & $9.24\substack{+0.09\\-0.10}$ & $29.82\substack{+9.38\\-4.40}$ & $1.91\substack{+0.36\\-0.48}$ & 1.90\\
	SPT2134-50 & 2.780 & $8.67\substack{+0.05\\-0.06}$ & $37.39\substack{+3.60\\-2.13}$ & $1.70\substack{+0.14\\-0.15}$ & 3.51\\
	SPT2146-55 & 4.567 & $9.03\substack{+0.10\\-0.10}$ & $33.32\substack{+8.78\\-5.89}$ & $1.80\substack{+0.39\\-0.39}$ & 1.01\\
	SPT2147-50 & 3.760 & $9.01\substack{+0.07\\-0.07}$ & $30.72\substack{+3.25\\-2.49}$ & $2.05\substack{+0.21\\-0.20}$ & 2.62\\
	SPT2152-40 & 3.851 & $9.29\substack{+0.22\\-0.11}$ & $30.11\substack{+6.51\\-9.32}$ & $1.92\substack{+0.69\\-0.38}$ & 4.47\\
	SPT2203-41 & 5.194 & $9.23\substack{+0.15\\-0.12}$ & $21.94\substack{+2.92\\-2.43}$ & $2.84\substack{+0.24\\-0.27}$ & 1.89\\
	SPT2232-61 & 2.894 & $9.18\substack{+0.06\\-0.06}$ & $27.91\substack{+2.27\\-2.59}$ & $2.24\substack{+0.19\\-0.15}$ & 2.93\\
	SPT2311-45 & 2.507 & $9.18\substack{+0.05\\-0.05}$ & $31.51\substack{+2.45\\-1.74}$ & $1.78\substack{+0.12\\-0.14}$ & 2.71\\
	SPT2311-54 & 4.280 & $9.18\substack{+0.06\\-0.06}$ & $40.77\substack{+3.49\\-4.39}$ & $1.83\substack{+0.20\\-0.16}$ & 4.16\\
	SPT2316-50 & 3.141 & $8.71\substack{+0.13\\-0.11}$ & $23.11\substack{+3.44\\-3.33}$ & $2.75\substack{+0.38\\-0.31}$ & 1.04\\
	SPT2319-55 & 5.293 & $8.74\substack{+0.10\\-0.10}$ & $29.34\substack{+6.12\\-3.78}$ & $2.21\substack{+0.29\\-0.33}$ & 1.10\\
	SPT2332-53 & 2.726 & $9.18\substack{+0.04\\-0.05}$ & $39.12\substack{+1.82\\-1.68}$ & $1.86\substack{+0.11\\-0.11}$ & 6.82\\
	SPT2335-53 & 4.756 & $8.46\substack{+0.11\\-0.09}$ & $30.55\substack{+4.68\\-5.40}$ & $2.68\substack{+0.38\\-0.31}$ & 1.93\\
	SPT2340-59 & 3.862 & $8.84\substack{+0.14\\-0.11}$ & $26.28\substack{+3.79\\-4.89}$ & $2.71\substack{+0.41\\-0.29}$ & 3.62\\
	SPT2349-50 & 2.876 & $9.37\substack{+0.07\\-0.07}$ & $30.69\substack{+3.06\\-4.05}$ & $2.06\substack{+0.24\\-0.18}$ & 3.26\\
	SPT2349-52 & 3.900 & $8.58\substack{+0.22\\-0.17}$ & $19.49\substack{+4.23\\-2.99}$ & $3.62\substack{+0.37\\-0.34}$ & 2.62\\
	SPT2351-57 & 5.811 & $8.65\substack{+0.07\\-0.06}$ & $39.58\substack{+4.69\\-4.72}$ & $1.98\substack{+0.23\\-0.21}$ & 3.76\\
	SPT2353-50 & 5.578 & $8.83\substack{+0.11\\-0.10}$ & $35.44\substack{+8.18\\-6.36}$ & $1.94\substack{+0.37\\-0.34}$ & 1.85\\
	SPT2354-58 & 1.867 & $9.09\substack{+0.05\\-0.07}$ & $31.95\substack{+3.77\\-2.00}$ & $2.05\substack{+0.12\\-0.16}$ & 1.73\\
	SPT2357-51 & 3.070 & $9.39\substack{+0.16\\-0.09}$ & $21.63\substack{+2.26\\-3.65}$ & $2.80\substack{+0.36\\-0.24}$ & 3.82\\
	\hline
\end{longtable}

\begin{longtable}{lccccr}
	
	\caption{Best-fitting SED parameters for all HerBS sources.}
	\label{tab:parameters_herbs}\\
	\hline
	\hline
	Source & Spec-z & log($M_{\textrm{dust}}$ [$M_\odot$]) & $T_{\textrm{dust}}$ [K] & $\beta$ & $\chi^2_\nu$ \\
	\hline
	HerBS-11 & 2.631 & $8.63\substack{+0.03\\-0.04}$ & $34.01\substack{+1.78\\-1.58}$ & $1.84\substack{+0.10\\-0.10}$ & 1.63\\
	HerBS-14 & 3.782 & $8.42\substack{+0.04\\-0.03}$ & $37.51\substack{+1.77\\-2.08}$ & $1.64\substack{+0.12\\-0.10}$ & 1.45\\
	HerBS-18 & 2.182 & $8.47\substack{+0.03\\-0.04}$ & $29.05\substack{+1.45\\-1.29}$ & $1.95\substack{+0.11\\-0.11}$ & 0.94\\
	HerBS-21 & 3.323 & $9.01\substack{+0.04\\-0.04}$ & $35.63\substack{+1.73\\-1.70}$ & $1.81\substack{+0.12\\-0.11}$ & 0.18\\
	HerBS-24 & 2.198 & $9.27\substack{+0.04\\-0.04}$ & $28.70\substack{+1.65\\-1.43}$ & $1.87\substack{+0.11\\-0.11}$ & 3.60\\
	HerBS-25 & 2.912 & $8.18\substack{+0.05\\-0.05}$ & $31.53\substack{+1.91\\-2.51}$ & $1.84\substack{+0.17\\-0.14}$ & 1.56\\
	HerBS-27 & 4.509 & $8.84\substack{+0.04\\-0.04}$ & $40.87\substack{+2.16\\-2.49}$ & $1.46\substack{+0.13\\-0.12}$ & 3.56\\
	HerBS-28 & 3.925 & $9.20\substack{+0.04\\-0.04}$ & $37.69\substack{+2.01\\-1.99}$ & $1.57\substack{+0.13\\-0.12}$ & 1.70\\
	HerBS-36 & 3.095 & $9.20\substack{+0.04\\-0.04}$ & $36.81\substack{+2.29\\-2.13}$ & $1.50\substack{+0.12\\-0.11}$ & 1.80\\
	HerBS-39 & 3.229 & $8.93\substack{+0.04\\-0.04}$ & $37.62\substack{+2.30\\-2.29}$ & $1.75\substack{+0.13\\-0.12}$ & 1.06\\
	HerBS-41 & 4.098 & $9.44\substack{+0.06\\-0.07}$ & $44.29\substack{+3.95\\-3.40}$ & $1.31\substack{+0.20\\-0.20}$ & 8.36\\
	HerBS-42 & 3.307 & $9.01\substack{+0.04\\-0.04}$ & $37.83\substack{+2.24\\-2.00}$ & $1.82\substack{+0.11\\-0.11}$ & 1.99\\
	HerBS-49 & 2.727 & $8.48\substack{+0.05\\-0.05}$ & $28.93\substack{+2.14\\-2.08}$ & $2.06\substack{+0.19\\-0.18}$ & 14.77\\
	HerBS-55 & 2.656 & $8.26\substack{+0.04\\-0.04}$ & $30.86\substack{+2.17\\-2.09}$ & $2.22\substack{+0.16\\-0.15}$ & 7.48\\
	HerBS-57 & 3.265 & $8.40\substack{+0.05\\-0.04}$ & $36.51\substack{+2.04\\-2.33}$ & $1.89\substack{+0.13\\-0.11}$ & 6.40\\
	HerBS-60 & 3.261 & $8.66\substack{+0.04\\-0.04}$ & $32.10\substack{+2.16\\-2.58}$ & $1.97\substack{+0.17\\-0.14}$ & 1.09\\
	HerBS-68 & 2.719 & $8.44\substack{+0.04\\-0.04}$ & $34.16\substack{+2.22\\-1.99}$ & $2.01\substack{+0.14\\-0.13}$ & 2.97\\
	HerBS-73 & 3.026 & $9.04\substack{+0.05\\-0.04}$ & $34.99\substack{+2.41\\-3.22}$ & $1.93\substack{+0.18\\-0.13}$ & 3.95\\
	HerBS-86 & 2.564 & $8.88\substack{+0.06\\-0.05}$ & $26.11\substack{+2.13\\-2.72}$ & $2.32\substack{+0.22\\-0.17}$ & 3.73\\
	HerBS-93 & 2.400 & $9.07\substack{+0.06\\-0.05}$ & $24.88\substack{+1.79\\-2.12}$ & $2.50\substack{+0.18\\-0.15}$ & 6.77\\
	HerBS-103 & 2.942 & $8.75\substack{+0.04\\-0.04}$ & $38.66\substack{+2.67\\-2.46}$ & $1.85\substack{+0.13\\-0.13}$ & 2.53\\
	HerBS-111 & 2.371 & $8.79\substack{+0.05\\-0.05}$ & $28.77\substack{+1.92\\-1.75}$ & $2.24\substack{+0.14\\-0.14}$ & 2.00\\
	HerBS-132 & 2.473 & $8.64\substack{+0.05\\-0.05}$ & $26.67\substack{+2.04\\-2.06}$ & $2.52\substack{+0.19\\-0.17}$ & 1.69\\
	HerBS-145 & 2.730 & $9.04\substack{+0.05\\-0.06}$ & $28.69\substack{+2.79\\-2.87}$ & $2.24\substack{+0.29\\-0.24}$ & 7.24\\
	HerBS-160 & 3.955 & $8.81\substack{+0.04\\-0.04}$ & $39.28\substack{+2.46\\-2.36}$ & $1.50\substack{+0.13\\-0.13}$ & 0.52\\
	HerBS-182 & 2.227 & $9.44\substack{+0.05\\-0.05}$ & $26.41\substack{+1.84\\-1.77}$ & $2.29\substack{+0.16\\-0.15}$ & 0.39\\
	HerBS-184 & 2.507 & $8.80\substack{+0.04\\-0.05}$ & $33.57\substack{+2.73\\-2.37}$ & $1.71\substack{+0.15\\-0.14}$ & 4.62\\
	HerBS-200 & 2.151 & $9.76\substack{+0.05\\-0.05}$ & $30.84\substack{+2.68\\-2.21}$ & $2.04\substack{+0.15\\-0.15}$ & 5.58\\
	HerBS-207 & 1.569 & $8.99\substack{+0.05\\-0.05}$ & $21.10\substack{+1.39\\-1.30}$ & $2.36\substack{+0.15\\-0.15}$ & 3.43\\
	HerBS-208 & 2.481 & $9.45\substack{+0.06\\-0.05}$ & $27.87\substack{+2.11\\-2.08}$ & $2.07\substack{+0.18\\-0.16}$ & 0.96\\
	\hline
\end{longtable}

\newpage

\section{Example Posterior Distributions}

\begin{figure*}
    \centering
    \includegraphics[width=0.5\textwidth]{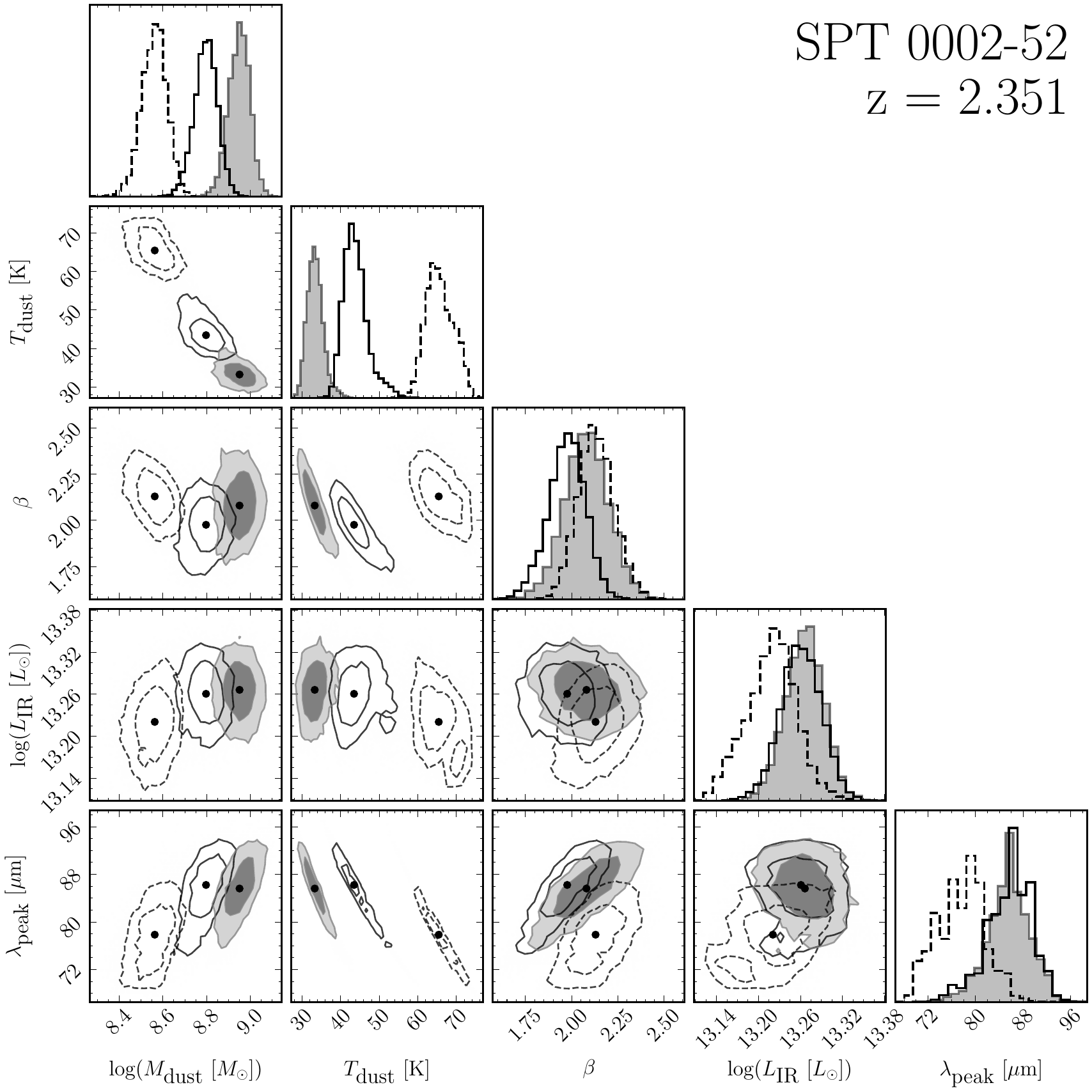}
    \includegraphics[width=0.5\textwidth]{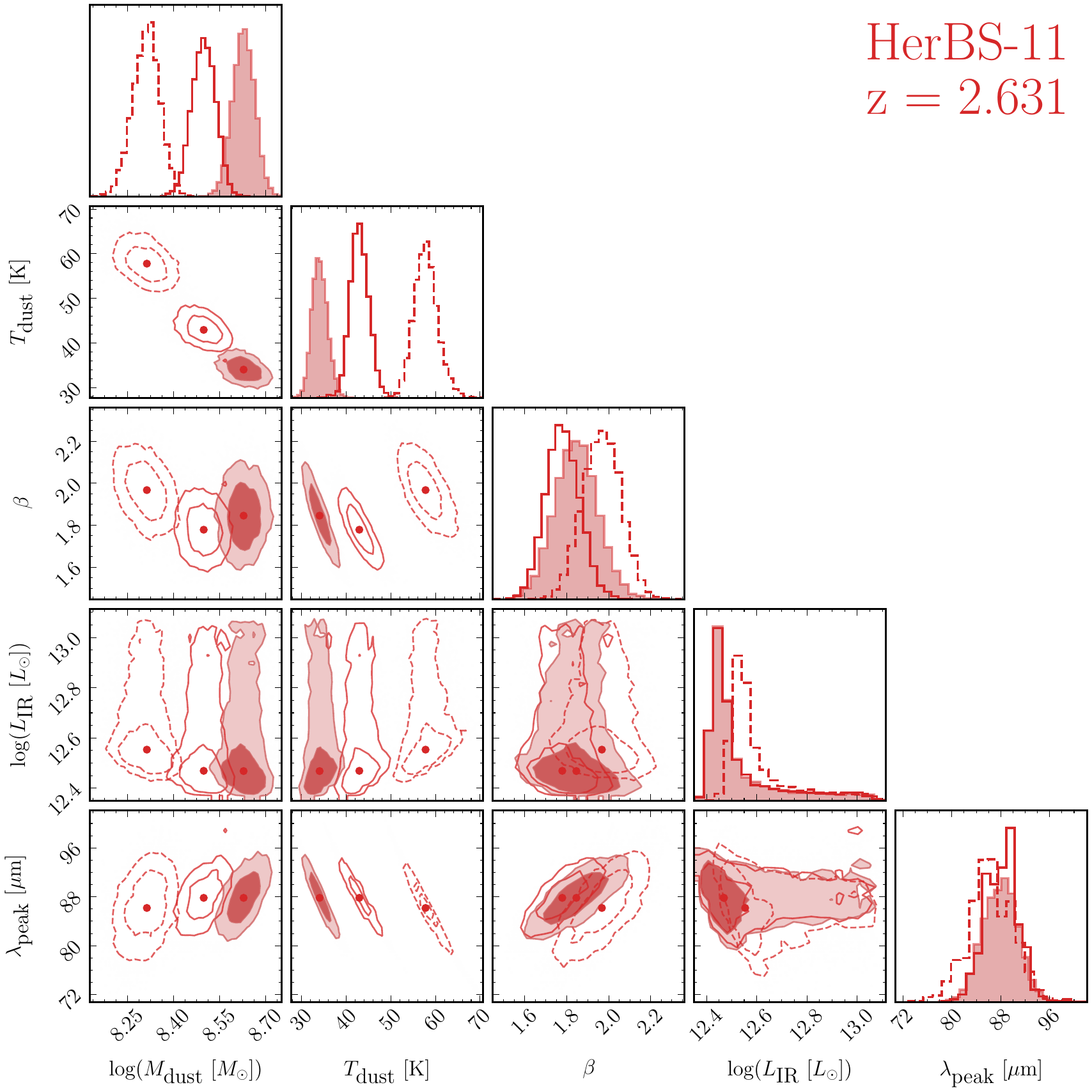}
    \caption{The posterior distributions for SPT 0002-52 and HerBS-11 obtained from FIR SED fitting with an optically thin dust model (shaded contours), and two general opacity models, one with $\lambda_1 = 100\,\mu$m (solid contours), the other with $\lambda_1 = 200\,\mu$m (dashed contours). Parameters shown are the dust mass $M_\textrm{dust}$, the dust temperature $T_\textrm{dust}$, the dust emissivity index $\beta$, IR luminosity $L_\textrm{IR}$ and the peak wavelength $\lambda_\textrm{peak}$.}
    \label{fig:corner_plots}
\end{figure*}

\label{lastpage}
\end{document}